\documentclass{article}%
\usepackage[top=3cm, bottom=3cm, left=3cm, right=3cm]{geometry}
\usepackage{amsmath}
\usepackage{amsfonts}
\usepackage{amssymb}
\usepackage{graphicx}
\usepackage[toc,page]{appendix}%
\providecommand{\U}[1]{\protect\rule{.1in}{.1in}}
\begin{document}

\title{Strong and radiative decays of excited vector mesons and predictions for a new
$\phi(1930)$ resonance}
\author{Milena Piotrowska$^{1}$, Christian Reisinger$^{2}$ and Francesco
Giacosa$^{1,2}$\\\textit{$^{1}$Institute of Physics, Jan Kochanowski University, }\\\textit{ul. Swietokrzyska 15, 25-406, Kielce, Poland.}\\\textit{$^{2}$Institute for Theoretical Physics, J. W. Goethe University, }\\\textit{ Max-von-Laue-Str. 1, 60438 Frankfurt, Germany.}}
\date{}
\maketitle

\begin{abstract}
We study the phenomenology of two nonets of excited vector mesons,
$\{\rho(1450)$, $K^{\ast}(1410)$, $\omega(1420)$, $\phi(1680)\}$ and
$\{\rho(1700),$ $K^{\ast}(1680),\omega(1650),$ $\phi(???)\},$ which (roughly)
correspond to radially excited $2^{3}S_{1}$ and to orbitally excited
$1^{3}D_{1}$ vector mesons. We evaluate the strong and radiative decays of
these mesons into pseudoscalar and ground-state vector mesons by using an
effective relativistic QFT model based on flavour symmetry. We compare decay
widths and branching ratios with various experimental results listed in the
PDG. An overall agreement of theory with experiment reinforces the standard
quark-antiquark assignment of the resonances mentioned above. Predictions for
not yet measured quantities are also made. In particular, we shall also make
predictions for the not-yet discovered $s\bar{s}$ state of the $1^{3}D_{1}$
nonet, denoted as $\phi(???)$. Its mass can be estimated to be about $1930$
MeV, hence we shall call this putative state $\phi(1930).$ Its main decays are
into $KK^{\ast}(892)$ (about $200$ MeV) and $KK$ (about $100$ MeV). Since this
state couples also to $\gamma\eta,$ it can be searched in the near future in
the photoproduction-based experiments GlueX and CLAS12 at Jefferson Lab.

\end{abstract}

\section{Introduction}

Strong interactions are described by Quantum Chromodynamics (QCD). While the
degrees of freedom of the QCD Lagrangian are elementary `colored' quarks and
gluons, the physical spectrum listed in the PDG \cite{pdg} consists of `white'
hadrons. Hadrons are, in general, bound states of quarks and gluons and are
further classified into mesons (bosonic hadrons) and baryons (fermionic hadrons).

The vast majority of mesons consists of `conventional' quark-antiquark states,
see the results of the quark model in\ Ref. \cite{godfrey}, but nowadays
experimental evidence for non-conventional mesons is mounting, e.g. Refs.
\cite{amsler,klempt}. Some known resonances in the low-energy sector might be
predominately four-quark state, e.g.\ Ref. \cite{pelaezrev}, or gluonic, e.g.
Ref. \cite{ochs}; in the energy region of charmonium and bottomonium masses
many resonances, the so-called $X,$ $Y$ and $Z$ states, have been found
\cite{xyzrev}. Similarly, conventional baryons are three-quark states
\cite{capstick}, but pentaquark states have been also recently discovered
\cite{pqlhcb}.

In this work we concentrate on the light mesonic sector by studying the
phenomenology (in particular, strong and radiative decays) of two types of
excited quark-antiquark vector mesons. The firm understanding of conventional
$\bar{q}q$ states is necessary to look for further non-conventional states
with the same quantum numbers. Moreover, the general status of excited states
is rather poorly understood and improvement is needed, see the quark model
review in the PDG \cite{qqpdg}. Excited vector mesons (for some previous
theoretical studies on them, see Refs.
\cite{afonin,caporale,badalian,coito1,coito2}) are especially interesting in
this context for various reasons:

(i) The ground-state vector mesons \{$\rho(770),$ $K^{\ast}(892),$
$\phi(1020),$ $\omega(782)$\} are very well known and represent an excellent
example of an ideal $\bar{q}q$ nonet. The non-relativistic quantum numbers are
$(n,L,S)=(1,0,1)$ ($n$ being the principal quantum number, $L$ the spacial
angular momentum, and $S$ the spin),\ hence, in the non-relativistic
spectroscopic notation one has $n$ $^{2S+1}L_{J}=1$ $^{3}S_{1}$ and in the
relativistic notation $J^{PC}=1^{--}.$ Considering that the nonet of
pseudoscalar meson is special due to the importance of spontaneous symmetry
breaking and the axial anomaly, one may say that ground-state vector mesons
are the lightest ideal $\bar{q}q$ objects.

(ii) Two types of excited vector mesons have been experimentally measured.
Although data need improvement, there is enough information on masses and
decays to undertake a systematic analysis. One nonet corresponds
(predominantly) to the radial excitation, with the quantum numbers
$(n,L,S)=(2,0,1)$ (spectroscopic notation $2^{3}S_{1},$ relativistic notation
$J^{PC}=1^{--});$ the associated states are \{$\rho(1450)$, $K^{\ast}(1410)$,
$\omega(1420)$, $\phi(1680)$\}. [Note, there are only two nonets of states
with $n=2$ listed in the PDG. Besides vector mesons, a tentative assignment
is done also for excited pseudoscalar mesons with $(n,L,S)=(2,0,0),$ for
details and references see the recent work in\ Ref. \cite{excitedscalars}.]
The second nonet corresponds (predominantly) to the orbital excitation, i.e.
it has quantum numbers $(n,L,S)=(1,2,1)$ (spectroscopic notation $1^{3}D_{1},$
relativistic notation still $J^{PC}=1^{--}$). The associated states are
\{$\rho(1700)$, $K^{\ast}(1680)$, $\omega(1650)$, $\phi(???)$\}. It should be
stressed from the very beginning that the physical resonances listed above do
not correspond exactly to the mentioned non-relativistic quark-antiquark
configurations, but can arise from mixing of them (moreover, a dressing of
meson-meson pairs also takes place). Hence we should regard the previous
assignment as indicating the predominant contribution.

(iii) The state $\phi(???)$ belonging to the nonet of orbitally excited
$1^{3}D_{1}$ states has not yet been observed. It is the last missing state of
that nonet, hence the empty space in Table II of Ref. \cite{qqpdg}. Its mass
can be estimated by the mass differences between the two nonets described
above, obtaining $m_{\phi(???)}-m_{\phi(1680)}\simeq$ $m_{\rho(1700)}%
-m_{\rho(1450)}\simeq250$ MeV, thus we expect that $\phi(???)$ has a mass of
about $1930$ MeV and we therefore postulate the existence of a new resonance,
denoted $\phi(1930)$ (note, the value $1930$ MeV is not far from the old quark
model prediction of $1880$ MeV \cite{godfrey}).

In this work, we study the decays of the excited vector mesons \{$\rho(1450)$,
$K^{\ast}(1410)$, $\omega(1420)$, $\phi(1680)$\} and \{$\rho(1700)$, $K^{\ast
}(1680)$, $\omega(1650)$, $\phi(???)\equiv\phi(1930)$\} by using a quantum
field theoretical (QFT) approach whose d.o.f. are mesons. The fields entering
in our model correspond to these resonances and hence roughly, but not
exactly, to the quark-antiquark configurations $2^{3}S_{1}$ and $1^{3}D_{1}$
predicted by the quark model (as discussed above mixing is possible). Indeed,
a direct link of our vector fields to an underlying microscopic wave function
of the quarks is not possible and also not necessary (details later on). The
Lagrangian of the model is constructed under the requirement of flavour
symmetry (i.e., the invariance under the rotation of the quarks $u,$ $d$, and
$s,$ a well proven approximate symmetry of QCD), as well as invariance under
parity $P$ and charge conjugation $C$. Indeed, flavour symmetry is present in
various approaches of low-energy QCD, such as chiral perturbation theory
\cite{chpt,chpt2}, linear sigma models
\cite{geffen,schechter,fariborz,dick,dick2}, and also Bethe-Salpeter
approaches \cite{bs,bs2}. Namely, even after spontaneous symmetry breaking
(SSB), flavour symmetry is still manifest. The use of flavour symmetric
mesonic QFT models, on which our approach is based, has proven to be
successful in the past in studies on decays of mesonic multiplets, such as the
well-known tensor mesons \cite{tensor}, the pseudovector mesons
\cite{pseudovector}, the pseudotensor mesons \cite{pseudotensor}, and also the
more difficult scalar mesons (in which the scalar glueball leaks in)
\cite{gutsche,gutsche2}. Here, we shall follow the very same methodology of
those works.

To be more specific, we write a Lagrangian that contains the two nonets of
excited vector mesons previously mentioned as well as their main decay
products: the ground-state vector mesons and pseudoscalar mesons. Moreover, in
a second step, we shall also include the photon for radiative decays. When
dominant terms in the large-$N_{c}$ expansion are kept, the number of
parameter of our Lagrangian is rather small: 4 parameters (2 for each nonet)
which allow to calculate a large number of decay rates (about 64 decay rates,
32 per each nonet). As a result, we can check the status of the
quark-antiquark assignment for these states and make predictions for decay
rates which are poorly known or have not yet been measured. In the PDG there
are also many branching ratios which can be compared to our results. We shall
find that the overall interpretation as $\bar{q}q$ states is satisfactory, but
some points deserve further clarification and more data is needed
(fortunately, the experiments GlueX and CLAS12 start operation soon). Then, as
a last important step, we will make predictions for the yet undiscovered
$\bar{s}s$ state $\phi(???)\equiv\phi(1930)$.

The paper is organized as follows: in\ Sec. 2 we present the fields of the
model, the Lagrangian, the theoretical expressions for the decay widths, as
well as the determination of the parameters. Then, in Sec. 3 we present the
results (separately for each excited nonet) of the decay widths and compare
them to the experimental results. In particular, we shall also discuss various
decay ratios. In Sec. 5 we summarize our results and present our outlook, in
particular in connection with the putative and yet undiscovered state
$\phi(1930)$. Various details are relegated to the appendices.

\section{The Model}

We use an effective relativistic Quantum Field Theoretical model based on
flavour symmetry. The d.o.f. are mesonic fields, which correspond to
quark-antiquark states. In this section, we first present the fields of the
model and their assignment, then we show the Lagrangian and finally the
expressions of the strong and radiative decay widths.

\subsection{The fields of the model}

As a first step, we introduce four nonets of mesons, which are given in terms
of matrices. Intuitively, each matrix corresponds to the following
quark-antiquark content:%

\begin{equation}
\frac{1}{\sqrt{2}}\left(
\begin{array}
[c]{lcr}%
\bar{u}u & \bar{d}u & \bar{s}u\\
\bar{u}d & \bar{d}d & \bar{s}d\\
\bar{u}s & \bar{d}s & \bar{s}s
\end{array}
\right)  \text{ .}%
\end{equation}
Although strictly speaking only the relativistic notation $J^{PC}$ is relevant
in a relativistic QFT treatment, we shall also keep track of the
non-relativistic notation for these multiplets, which should be heuristically
understood as the dominant contribution in the (in our approach invisible)
microscopic wave function of the quark and the antiquark pairs. In this way, a
link with the quark model results -although approximate- is possible and
allows for a better intuitive understanding of these states. The explicit form
of the matrices for the nonet of pseudoscalar mesons, the nonet of
ground-state vector mesons, and two nonets of excited vector mesons read:
\begin{equation}
P=\frac{1}{\sqrt{2}}\left(
\begin{array}
[c]{lcr}%
\frac{\eta_{N}+\pi^{0}}{\sqrt{2}} & \pi^{+} & K^{+}\\
\pi^{-} & \frac{\eta_{N}-\pi^{0}}{\sqrt{2}} & K^{0}\\
K^{-} & \bar{K}^{0} & \eta_{S}%
\end{array}
\right)  \text{ ; }V^{\mu}=\frac{1}{\sqrt{2}}\left(
\begin{array}
[c]{lcr}%
\frac{\omega^{\mu}+\rho^{\mu0}}{\sqrt{2}} & \rho^{\mu+} & K_{i}^{\mu\star+}\\
\rho^{\mu-} & \frac{\omega^{\mu}-\rho^{\mu0}}{\sqrt{2}} & K^{\mu\star0}\\
K^{\mu\star-} & \bar{K}^{\mu\star0} & \phi^{\mu}%
\end{array}
\right)  \text{ ; } \label{pv}%
\end{equation}

\begin{equation}
V_{E}^{\mu}=\frac{1}{\sqrt{2}}\left(
\begin{array}
[c]{lcr}%
\frac{\omega_{E}^{\mu}+\rho_{E}^{\mu0}}{\sqrt{2}} & \rho_{E}^{\mu+} &
K_{E}^{\mu\star+}\\
\rho_{E}^{\mu-} & \frac{\omega_{E}^{\mu}-\rho_{E}^{\mu0}}{\sqrt{2}} &
K_{E}^{\mu\star0}\\
K_{E}^{\mu\star-} & \bar{K}_{E}^{\mu\star0} & \phi_{E}^{\mu}%
\end{array}
\right)  \text{ ; }V_{D}^{\mu}=\frac{1}{\sqrt{2}}\left(
\begin{array}
[c]{lcr}%
\frac{\omega_{D}^{\mu}+\rho_{D}^{\mu0}}{\sqrt{2}} & \rho_{D}^{\mu+} &
K_{D}^{\mu\star+}\\
\rho_{D}^{\mu-} & \frac{\omega_{D}^{\mu}-\rho_{D}^{\mu0}}{\sqrt{2}} &
K_{D}^{\mu\star0}\\
K_{D}^{\mu\star-} & \bar{K}_{D}^{\mu\star0} & \phi_{D}^{\mu}%
\end{array}
\right)  \text{ .} \label{ve}%
\end{equation}
(For the explicit quark-antiquark microscopic current leading to these fields,
see Ref. \cite{pseudotensor}). The matrix $P$ describes the nonet of
pseudoscalar mesons corresponding to the states $\{\pi,K,\eta\equiv
\eta(547),\eta^{\prime}\equiv\eta^{\prime}(958)\}.$ Namely, the two fields
denoted as $\eta_{N}$ and $\eta_{S}$ in\ Eq. (\ref{pv}) mix and generate the
physical fields $\eta$ and $\eta^{\prime}$:
\begin{equation}
\eta_{N}=\eta\cos\theta_{P}-\eta^{\prime}\sin\theta_{P}\text{ and }\eta
_{S}=\eta\sin\theta_{P}-\eta^{\prime}\cos\theta_{P}\text{ .}%
\end{equation}
The mixing angle $\theta_{P}$ is set to $-42^{\circ}$ \cite{kloe2}. Using
other values in the range ($-40^{\circ},$ $-45^{\circ}$), e.g. Refs.
\cite{dick,feldmann,bass}, would affect only marginally our results. In the
non-relativistic notation, this nonet corresponds (predominantly) to
$(n,L,S)=(1,0,0),$ hence to $n$ $^{2S+1}L_{J}=1$ $^{1}S_{0}$. The relativistic
notation is $J^{PC}=0^{-+}.$

The nonet of ground-state vector mesons, denoted as $V^{\mu}$, is associated
to the resonances $\{\rho(770),$ $K^{\ast}(892),$ $\phi(1020),$ $\omega
(782)\}$. Actually, also the bare fields $\omega=\sqrt{1/2}(\bar{u}u+\bar
{d}d)$ and $\phi=\bar{s}s$ entering in Eq. (\ref{pv}) mix, but the mixing is
sufficiently small to be neglected (about $-3^{\circ},$ \cite{qqpdg}); then,
$\omega(782)$ is regarded as purely nonstrange and $\phi(1020)$ as purely
strange. The nonrelativistic quantum numbers are $(n,L,S)=(1,0,1),$ hence $n$
$^{2S+1}L_{J}=1$ $^{3}S_{1}$. Relativistically: $J^{PC}=1^{--}.$

Finally, we turn to the excited vector mesons. The matrix $V_{E}^{\mu}$
describes the first nonet of excited vector mesons that we assign to the
states \{$\rho(1450)$, $K^{\ast}(1410),$ $\omega(1420),$ $\phi(1680)$\}. As
for vector mesons, we neglect the isoscalar mixing, hence $\omega(1420)$ is
purely nonstrange ($\sqrt{1/2}(\bar{u}u+\bar{d}d)$ ) and $\phi(1680)$ purely
strange, $\bar{s}s$. This nonet corresponds to (predominantly but not
identically) radially excited vector mesons with the non-relativistic quantum
numbers $(n,L,S)=(2,0,1),$ hence $n$ $^{2S+1}L_{J}=2$ $^{3}S_{1}$.
Relativistically: $J^{PC}=1^{--}$ (just as the ground-state vector mesons).
Note, the state $K^{\ast}(1410)$ has been clearly seen in the recent lattice
studies of Ref. \cite{prelovsek} while $\omega(1420)$ and $\phi(1680)$ have
been observed in the lattice studies of Ref. \cite{dudek,dudek2}.

The matrix $V_{D}^{\mu}$ describes the second nonet of excited vectors:
\{$\rho(1700)$, $K^{\ast}(1680)$, $\omega(1650)$, $\phi(???)$\}. Also here,
the isoscalar mixing is neglected. Non relativistically, this nonet
corresponds (always predominantly) to orbitally excited vector mesons with
$(n,L,S)=(1,2,1),$ hence $n$ $^{2S+1}L_{J}=1$ $^{3}D_{1}$. Relativistically:
$J^{PC}=1^{--}.$ The $\bar{s}s$ state in the $1^{3}D_{1}$ nonet, denoted as
$\phi(???)$, has not yet been experimentally seen. By using our model, we make
predictions for this resonance. In order to include this state into our
calculation, we have to estimate its mass. In Table 1 we present the masses of
the known members of the two nonets. It is visible that the mass difference
between radially and orbitally excited vector mesons are approximately the
same for all states. The reason for that is the same type of strong dynamics
describing these mesons and their masses.

\begin{table}[h]
\makebox[\textwidth][c] {%
\begin{tabular}
[c]{|c|c|c|c|c|}\hline
$V_{E}$ & $\rho(1450)$ & $K^{\ast}(1410)$ & $\omega(1420)$ & $\phi
(1680)$\\\hline
$V_{D}$ & $\rho(1700)$ & $K^{\ast}(1680)$ & $\omega(1650)$ & $\phi
(???)$\\\hline
Difference & $250$ MeV & $270$ MeV & $230$ MeV & ?\\\hline
\end{tabular}
}\caption{Mass differences between the members of the two nonets of excited
vector mesons.}%
\end{table}Hence, we can estimate the mass of $\phi(???)$ as:
\begin{equation}
m_{\phi(???)}\simeq(m_{\phi(1680)}+250\pm20)\text{ MeV}=1930\pm20\text{ MeV}.
\end{equation}
From now on we shall call this hypothetical state
\[
\phi(???)\equiv\phi(1930)\text{ .}%
\]

All the fields have precise transformations under parity $P,$ charge
conjugation $C,$ and flavour symmetry $U(3)_{V},$ which are summarized in
Table 2.

\begin{table}[h]
\makebox[\textwidth][c] {%
\begin{tabular}
[c]{|c|c|c|c|}\hline
& Parity ($P$) & Charge conjugation ($C$) & Flavor ($U\left(  3\right)  _{V}%
$)\\\hline\hline
$P$ & $-P(t,-\vec{x})$ & $P^{t}$ & $UPU^{\dagger}$\\\hline
$V^{\mu}$ & $V_{\mu}(t,-\vec{x})$ & $-\left(  V^{\mu}\right)  ^{t}$ &
$UV^{\mu}U^{\dagger}$\\\hline
$V_{E}^{\mu}$ & $V_{E,\mu}(t,-\vec{x})$ & $-\left(  V_{E}^{\mu}\right)  ^{t}$
& $UV_{E}^{\mu}U^{\dagger}$\\\hline
$V_{D}^{\mu}$ & $V_{D,\mu}(t,-\vec{x})$ & $-\left(  V_{D}^{\mu}\right)  ^{t}$
& $UV_{D}^{\mu}U^{\dagger}$\\\hline
\end{tabular}
}\caption{Transformation properties of the nonets under charge, parity, and
flavour transformations. Notice that the parity transformation for vector
states is obtained by lowering the Lorentz index. }%
\end{table}

\subsection{The Lagrangian}

The Lagrangian of the model is obtained by properly coupling the matrices
listed above and by requiring invariance under $P,$ $C,$ and $U(3)_{V}.$ It
explicitly reads:
\begin{equation}
\mathcal{L}=\mathcal{L}_{EPP}+\mathcal{L}_{DPP}+\mathcal{L}_{EVP}%
+\mathcal{L}_{DVP} \label{lagfull}%
\end{equation}
where%
\begin{align}
\mathcal{L}_{EPP}  &  =ig_{EPP}Tr\left(  [\partial^{\mu}P,V_{E,\mu}]P\right)
\text{ ,\quad}\mathcal{L}_{DPP}=ig_{DPP}Tr\left(  [\partial^{\mu}P,V_{D,\mu
}]P\right)  \text{ ,}\label{lagfull1}\\
\mathcal{L}_{EVP}  &  =g_{EVP}Tr\left(  \tilde{V}_{E}^{\mu\nu}\{V_{\mu\nu
},P\}\right)  \text{ ,}\quad\mathcal{L}_{DVP}=g_{DVP}Tr\left(  \tilde{V}%
_{D}^{\mu\nu}\{V_{\mu\nu},P\}\right)  \text{ .} \label{lagfull2}%
\end{align}
The terms contain various processes: $\mathcal{L}_{EPP}$ describes the decay
$V_{E}\rightarrow PP,$ $\mathcal{L}_{DPP}$ the decay $V_{D}\rightarrow PP,$
$\mathcal{L}_{EVP}$ the decay $V_{E}\rightarrow VP$, and finally
$\mathcal{L}_{DVP}$ the decay $V_{D}\rightarrow VP.$ The notation
$[A,B]=AB-BA$ stands for the usual commutator and $\{A,B\}=AB+BA$ for the
anticommutator. Moreover, the dual fields have been defined in the standard
way:
\begin{equation}
\tilde{V}_{E}^{\mu\nu}=\frac{1}{2}\epsilon^{\mu\nu\alpha\beta}(\partial
_{\alpha}V_{E,\beta}-\partial_{\beta}V_{E,\alpha})\text{ ,}%
\end{equation}%
\begin{equation}
\tilde{V}_{D}^{\mu\nu}=\frac{1}{2}\epsilon^{\mu\nu\alpha\beta}(\partial
_{\alpha}V_{D,\beta}-\partial_{\beta}V_{D,\alpha})\text{ .}%
\end{equation}
For every term of the Lagrangian there is a corresponding coupling constant:
$g_{EPP},$ $g_{DPP},$ $g_{EVP},$ $g_{DVP}$, hence the model contains four
parameters. To fix them we use some of the experimental data taken from PDG,
see Sec. 3.1. The extended forms of the interaction terms are presented in
Appendix A.

Finally, we shall also study the radiative decays of the type $V\rightarrow
\gamma P$. To this end, we need to perform the following replacement of the
vector field strength tensor as (see e.g. Ref. \cite{connell}):
\begin{equation}
V_{\mu\nu}\rightarrow V_{\mu\nu}+\frac{e_{0}}{g_{\rho}}QF_{\mu\nu}\text{ ,}
\label{gammasubst}%
\end{equation}
where $F_{\mu\nu}$ is the field strength tensor for photons, $e_{0}=\sqrt
{4\pi\alpha}$ (with $\alpha\approx1/137$) is the electric charge of the
proton, $g_{\rho}=5.5\pm0.5$ is the $\rho\pi\pi$ coupling constant, and
$Q=diag\{2/3,-1/3,-1/3\}$ is the matrix with the charges of the quarks. Note,
radiative decays do not necessitate any new parameter. For more details, see
Appendix B.

Next, we discuss three important theoretical aspects and further
developments/improvements of our model.

(i) Large-$N_{c}$ suppressed terms. In our framework, all the nonets are
interpreted as $\bar{q}q$ states, hence the coupling constants $g_{EPP}%
,g_{DPP},g_{EVP},g_{DVP}$ scale as $1/\sqrt{N_{c}}$ and are dominant in the
large-$N_{c}$ expansion \cite{thooft} (for a review, see Ref. \cite{witten}).
For instance, for what concerns the decay, $\tilde{V}_{E}^{\mu\nu}\rightarrow
VP$ further flavour-symmetric terms which are suppressed in the large-$N_{c}$
limit have the form
\begin{equation}
Tr\left(  \tilde{V}_{E}^{\mu\nu}\right)  Tr\left(  V_{\mu\nu}P\right)  \text{
and }Tr\left(  \tilde{V}_{E}^{\mu\nu}\right)  Tr\left(  V_{\mu\nu}\right)
Tr\left(  P\right)
\end{equation}
(similarly, for $\tilde{V}_{D}^{\mu\nu}\rightarrow VP$). The corresponding
coupling constants are proportional to $1/N_{c}^{3/2}$ and $1/N_{c}^{5/2}$,
respectively. At the present level of accuracy of the experimental data, these
terms can be safely neglected. Once more precise data will be available, they
can be included. Note, such terms do not exist for the $V_{E}\rightarrow PP$
and $V_{D}\rightarrow PP$ because of the anticommutator (in turn, this is the
reason why $\omega_{E,D}\rightarrow\pi\pi$ and $\phi_{E,D}\rightarrow\pi\pi$ vanish).

(ii) Flavour breaking terms. There are terms which break explicitly flavour
symmetry, such as%
\begin{equation}
iTr\left(  \tilde{\lambda}[\partial^{\mu}P,V_{E,\mu}]P\right)  \text{ ,}%
\end{equation}
where $\tilde{\lambda}\propto diag\{0,m_{d}-m_{u},m_{s}-m_{u}\}$ is
proportional to the mass differences. Typically, $m_{d}-m_{u}$ can be safely
neglected, but $m_{s}-m_{u}$ can be non-negligible (in Refs.
\cite{gutsche,gutsche2} there is a contribution of about $10$ $\%$ from such a
term). Also in this case, these effects are not taken into account here
because the precision of data would not allow to constrain them. Note, one
could also write down terms which break flavour symmetry and are subleading in
the large-$N_{c}$ expansions, but they are doubly suppressed and neglected in
the present work.

(iii) Mixing between excited vectors. Within our treatment, the mesonic fields
correspond directly to the physical fields. Indeed, one could start with two
nonets of fields $V_{E,bare}^{\mu}$ and $V_{D,bare}^{\mu}$ which correspond to
purely radially excited ($2^{3}S_{1}$) and orbitally excited ($1^{3}D_{1}$)
mesons, respectively. In this case, one would write down a Lagrangian
analogous to the one of Eqs. (\ref{lagfull1})-(\ref{lagfull2}), where however
the coupling should be named differently: $g_{EPP}^{bare},g_{DPP}%
^{bare},g_{EVP}^{bare},g_{DVP}^{bare}$. In addition, one should add the mixing
term between the bare configurations:
\begin{equation}
\delta_{mix}Tr[V_{E,bare,\mu}V_{D,bare}^{\mu}]\text{ .}%
\end{equation}
Then, one should perform the usual $O(2)$ rotation:
\begin{align}
V_{E,\mu}  &  =V_{E,bare,\mu}\cos\theta_{ED}+V_{D,bare}^{\mu}\sin\theta
_{ED}\text{ ,}\\
V_{D,\mu}  &  =-V_{E,bare,\mu}\sin\theta_{ED}+V_{D,bare}^{\mu}\cos\theta
_{ED}\text{ .}%
\end{align}
The previous equations are valid in the flavour limit because all the members
of the nonet rotate with the same mixing angle. In terms of the physical
$V_{E,\mu}$ and $V_{D,\mu}$ no mixing is present. For instance, the coupling
constants $g_{EPP}$ reads in terms of the bare couplings as:
\begin{equation}
g_{EPP}=g_{EPP}^{bare}\cos\theta_{ED}+g_{EPP}^{bare}\sin\theta_{ED}\text{ ;}%
\end{equation}
similar relations hold for the other coupling constants. The important point
is that, once we have performed the rotation, the fields $V_{E}$ and $V_{D}$
correspond to the physical ones and the mixing angle $\theta_{ED}$ cannot be
determined as long as flavour symmetry is valid (in fact, it completely
disappears from all physical quantities). This is why, by no loss of
generality, we did not include any mixing term in our Lagrangian of Eqs.
(\ref{lagfull1})-(\ref{lagfull2}). Thus, while we cannot state that the
physical fields have a certain microscopic wave function, our analysis is more
general in the sense that the fields of our model are already a mixture of the
bare quark-model configurations. Even we do not expect $\theta_{ED}$ to be
large, our analysis is independent on its precise value. Indeed, the only way
to render the mixing of bare configurations visible is to include violations
of flavour symmetry discussed above in point (ii). In that way, different
mixing angles would emerge and one could not \textquotedblleft rotate
away\textquotedblright\ the mixing. Such deviations are anyhow expected to be
small, as the splitting of the masses in Table 2 shows (similar mass
differences between the corresponding members of the multiplets) .

In conclusion, the (here neglected) effects (i), (ii), and (iii) show how to
potentially improve the present model in a systematic way. This is left as a
work for the future.

\subsection{Decay widths: theoretical expressions}

The tree-level decay widths for a resonance $R=V_{E}\equiv E$ or
$R=V_{D}\equiv D$ can be calculated by performing a standard QFT calculation.
The results for the three channels that we consider (pseudoscalar-pseudoscalar
($PP$), vector-pseudoscalar ($VP$), and photon-pseudoscalar ($\gamma P$)) read
explicitly:
\begin{equation}
\Gamma_{R\rightarrow PP}=s_{RPP}\frac{|\vec{k}|^{3}}{6\pi m_{R}^{2}}\left(
\frac{g_{RPP}}{2}\lambda_{RPP}\right)  ^{2}\text{ ,} \label{rpp}%
\end{equation}%
\begin{equation}
\Gamma_{R\rightarrow VP}=s_{RVP}\frac{|\vec{k}|^{3}}{12\pi}\left(
\frac{g_{RVP}}{2}\lambda_{RVP}\right)  ^{2}\text{ ,} \label{rvp}%
\end{equation}%
\begin{equation}
\Gamma_{R\rightarrow\gamma P}=\frac{|\vec{k}|^{3}}{12\pi}\left(  \frac
{g_{RVP}}{2}\frac{e_{0}}{g_{\rho}}\lambda_{R\gamma P}\right)  ^{2}\text{ ,}
\label{rgammap}%
\end{equation}
where\newline%
\begin{equation}
|\vec{k}|=\frac{\sqrt{m_{R}^{4}+(m_{A}^{2}-m_{B}^{2})^{2}-2(m_{A}^{2}%
+m_{B}^{2})m_{R}^{2}}}{2m_{R}}%
\end{equation}
is the modulus of the three-momentum of the outgoing particles. Moreover,
$m_{R}$ refers to the mass of the decaying resonance, while $m_{A}$ and
$m_{B}$ to the masses of decay products. In Tables 3, 4, and 5 we report the
flavour degeneracy coefficients $s_{RPP},$ $s_{RVP}$ as well as the
Clebsch-Gordon coefficients $\lambda_{RPP},$ $\lambda_{RVP},$ $\lambda
_{R\gamma P}$ arising from the explicit evaluation of traces (see Appendix A).

\begin{table}[h]
\renewcommand{\arraystretch}{1.23}
\par
\makebox[\textwidth][c] { 
\par%
\begin{tabular}
[c]{|c|c|c|c|}\hline
\multicolumn{2}{|c|}{Decay channel} & Symmetry factor, Eq. (\ref{rpp}) &
Amplitude, Eq. (\ref{rpp})\\\cline{1-2}%
$V_{E}\rightarrow PP$ & $V_{D}\rightarrow PP$ & $s_{EPP}=s_{DPP}$ &
$\lambda_{EPP}=\lambda_{DPP}$\\\hline\hline
$\rho(1450)\rightarrow\bar{K}K$ & $\rho(1700)\rightarrow\bar{K}K$ & $2$ &
$\frac{1}{2}$\\\hline
$\rho(1450)\rightarrow\pi\pi$ & $\rho(1700)\rightarrow\pi\pi$ & $1$ &
$1$\\\hline
$K^{\ast}(1410)\rightarrow K\pi$ & $K^{\ast}(1680)\rightarrow K\pi$ & $3$ &
$\frac{1}{2}$\\\hline
$K^{\ast}(1410)\rightarrow K\eta$ & $K^{\ast}(1680)\rightarrow K\eta$ & $1$ &
$\frac{1}{2}(\cos\theta_{p}-\sqrt{2}\sin\theta_{p})$\\\hline
$K^{\ast}(1410)\rightarrow K\eta^{\prime}$ & $K^{\ast}(1680)\rightarrow
K\eta^{\prime}$ & $1$ & $\frac{1}{2}(\sqrt{2}\cos\theta_{p}+\sin\theta_{p}%
)$\\\hline
$\omega(1420)\rightarrow\bar{K}K$ & $\omega(1650)\rightarrow\bar{K}K$ & $2$ &
$\frac{1}{2}$\\\hline
$\phi(1680)\rightarrow\bar{K}K$ & $\phi(1930)\rightarrow\bar{K}K$ & $2$ &
$\frac{1}{\sqrt{2}}$\\\hline
\end{tabular}
}\caption{All symmetry factors and the amplitude's coefficents of Eq.
(\ref{rpp}). They can be extracted from\ the Lagrangian of Eq. (\ref{lagfull1}%
), whose expanded form is presented in\ Appendix A.}%
\end{table}

\begin{table}[h]
\renewcommand{\arraystretch}{1.23}
\par
\makebox[\textwidth][c] {
\begin{tabular}
[c]{|c|c|c|c|}\hline
\multicolumn{2}{|c|}{Decay channel} & Symmetry factor, Eq. (\ref{rvp}) &
Amplitude, Eq. (\ref{rvp})\\\cline{1-2}%
$V_{E}\rightarrow VP$ & $V_{D}\rightarrow VP$ & $s_{EVP}=s_{DVP}$ &
$\lambda_{EVP}=\lambda s_{DVP}$\\\hline\hline
$\rho(1450)\rightarrow\omega\pi$ & $\rho(1700)\rightarrow\omega\pi$ & $1$ &
$\frac{1}{2}$\\\hline
$\rho(1450)\rightarrow K^{\ast}(892)K$ & $\rho(1700)\rightarrow K^{\ast
}(892)K$ & $4$ & $\frac{1}{4}$\\\hline
$\rho(1450)\rightarrow\rho(770)\eta$ & $\rho(1700)\rightarrow\rho(770)\eta$ &
$1$ & $\frac{1}{2}\cos\theta_{p}$\\\hline
$\rho(1450)\rightarrow\rho(770)\eta^{\prime}$ & $\rho(1700)\rightarrow
\rho(770)\eta^{\prime}$ & $1$ & $\frac{1}{2}\sin\theta_{p}$\\\hline
$K^{\ast}(1410)\rightarrow K\rho$ & $K^{\ast}(1680)\rightarrow K\rho$ & $3$ &
$\frac{1}{4}$\\\hline
$K^{\ast}(1410)\rightarrow K\phi$ & $K^{\ast}(1680)\rightarrow K\phi$ & $1$ &
$\frac{1}{2\sqrt{2}}$\\\hline
$K^{\ast}(1410)\rightarrow K\omega$ & $K^{\ast}(1680)\rightarrow K\omega$ &
$1$ & $\frac{1}{4}$\\\hline
$K^{\ast}(1410)\rightarrow K^{\ast}(892)\pi$ & $K^{\ast}(1680)\rightarrow
K^{\ast}(892)\pi$ & $3$ & $\frac{1}{4}$\\\hline
$K^{\ast}(1410)\rightarrow K^{\ast}(892)\eta$ & $K^{\ast}(1680)\rightarrow
K^{\ast}(892)\eta$ & $1$ & $\frac{1}{4}(\cos\theta_{p}+\sqrt{2}\sin\theta
_{p})$\\\hline
$K^{\ast}(1410)\rightarrow K^{\ast}(892)\eta^{\prime}$ & $K^{\ast
}(1680)\rightarrow K^{\ast}(892)\eta^{\prime}$ & $2$ & $\frac{1}{4}(\sqrt
{2}\cos\theta_{p}-\sin\theta_{p})$\\\hline
$\omega(1420)\rightarrow\rho\pi$ & $\omega(1650)\rightarrow\rho\pi$ & $3$ &
$\frac{1}{2}$\\\hline
$\omega(1420)\rightarrow K^{\ast}(892)K$ & $\omega(1650)\rightarrow K^{\ast
}(892)K$ & $4$ & $\frac{1}{4}$\\\hline
$\omega(1420)\rightarrow\omega(782)\eta$ & $\omega(1650)\rightarrow
\omega(782)\eta$ & $1$ & $\frac{1}{2}\cos\theta_{p}$\\\hline
$\omega(1420)\rightarrow\omega(782)\eta^{\prime}$ & $\omega(1650)\rightarrow
\omega(782)\eta^{\prime}$ & $1$ & $\frac{1}{2}\sin\theta_{p}$\\\hline
$\phi(1680)\rightarrow K\bar{K}^{\ast}$ & $\phi(1930)\rightarrow K\bar
{K}^{\ast}$ & $4$ & $\frac{1}{2\sqrt{2}}$\\\hline
$\phi(1680)\rightarrow\phi(1020)\eta$ & $\phi(1930)\rightarrow\phi(1020)\eta$
& $1$ & $\frac{1}{\sqrt{2}}\sin\theta_{p}$\\\hline
$\phi(1680)\rightarrow\phi(1020)\eta^{\prime}$ & $\phi(1930)\rightarrow
\phi(1020)\eta^{\prime}$ & $1$ & $\frac{1}{\sqrt{2}}\cos\theta_{p}$\\\hline
\end{tabular}
}\caption{All symmetry factors and the amplitude's coefficents of Eq.
(\ref{rvp}). They can be extracted from\ the Lagrangian of Eq. (\ref{lagfull2}%
), whose expanded form is presented in\ Appendix A.}%
\end{table}

\begin{table}[h]
\renewcommand{\arraystretch}{1.23}
\par
\makebox[\textwidth][c] {%
\begin{tabular}
[c]{|c|c|c|}\hline
\multicolumn{2}{|c|}{Decay channel} & Amplitude, Eq. (\ref{rgammap}%
)\\\cline{1-2}%
$V_{E}\rightarrow\gamma P$ & $V_{D}\rightarrow\gamma P$ & $\lambda_{E\gamma
P}=\lambda_{D\gamma P}$\\\hline\hline
$\rho(1450)\rightarrow\gamma\pi$ & $\rho(1700)\rightarrow\gamma\pi$ &
$\frac{1}{6}$\\\hline
$\rho(1450)\rightarrow\gamma\eta$ & $\rho(1700)\rightarrow\gamma\eta$ &
$\frac{1}{2}\cos\theta_{p}$\\\hline
$\rho(1450)\rightarrow\gamma\eta^{\prime}$ & $\rho(1700)\rightarrow\gamma
\eta^{\prime}$ & $\frac{1}{2}\sin\theta_{p}$\\\hline
$K^{\ast}(1410)\rightarrow\gamma K$ & $K^{\ast}(1680)\rightarrow\gamma K$ &
$\frac{1}{3}$\\\hline
$\omega(1420)\rightarrow\gamma\pi$ & $\omega(1650)\rightarrow\gamma\pi$ &
$\frac{1}{2}$\\\hline
$\omega(1420)\rightarrow\gamma\eta$ & $\omega(1650)\rightarrow\gamma\eta$ &
$\frac{1}{6}\cos\theta_{p}$\\\hline
$\omega(1420)\rightarrow\gamma\eta^{\prime}$ & $\omega(1650)\rightarrow
\gamma\eta^{\prime}$ & $\frac{1}{6}\cos\theta_{p}$\\\hline
$\phi(1680)\rightarrow\gamma\eta$ & $\phi(1930)\rightarrow\gamma\eta$ &
$\frac{1}{3}\sin\theta_{p}$\\\hline
$\phi(1680)\rightarrow\gamma\eta^{\prime}$ & $\phi(1930)\rightarrow\gamma
\eta^{\prime}$ & $\frac{1}{3}\cos\theta_{p}$\\\hline
\end{tabular}
}\caption{Amplitude's coefficients of Eq. (\ref{rgammap}) extracted from Eq.
(\ref{lagfull2}) together with the shift of Eq. (\ref{gammasubst}).}%
\end{table}

The inclusion of loops and the evaluation of the positions of the poles would
allow to go beyond tree-level; this is also left as an outlook. For the decays
that we will examine, the ratio `width'/`mass' ($\Gamma/M$) is safely below 1,
ensuring that loop corrections do not change much the tree-level results
\cite{lupo}. Yet, the study of the pole trajectories and the eventual
generation of additional poles (a typical QFT phenomenon that occurs when the
coupling strength is strong enough, e.g. Refs.
\cite{pennington,wolkanowski,milenathomas,milenaacta}) are surely interesting
and could be addressed within our framework at a later stage.

\section{Results}

In this section we present the results. First, in Sec. 3.1 we determine the
parameters of the model by using some selected data from PDG. Then, in Sec.
3.2 we concentrate on the results for the decays of the states \{$\rho
(1450),K^{\ast}(1410),\omega(1420),\phi(1680)$\} and in Sec. 3.3 for the
states \{$\rho(1700)$, $K^{\ast}(1680)$, $\omega(1650)$, $\phi(???)\equiv
\phi(1930)$\}. In both cases we shall present summarizing tables and compare
to additional ratios quoted in the PDG. When referring to a particular
experiment, we shall use the notation of the PDG (first author and year in
which the corresponding publication appeared).

\subsection{Determination of the coupling constants}

In order to determine the coupling constants one has to choose some well-known
four experimental values (two for each nonet). At present a full fit to all
experimental values does not seem the best procedure. In some cases, some
observables were measured by a single experiment, in other cases different
experimental results are not compatible with each other. Hence, the choice of
4 rather stable experimental results seems the best strategy to fix our four
parameters. Later on, it will be possible to compare the results to partial
widths and to quite many ratios between partial widths, see Secs. 3.2 and 3.3.

For what concerns $g_{EPP}$ and $g_{EVP},$ we use the following experimental
data taken from PDG \cite{pdg}:
\begin{align}
\Gamma_{K^{\ast}(1410)\rightarrow K\pi}^{exp}  &  =15.3\pm3.3\text{ MeV}\\
\Gamma_{\phi(1680)}^{tot,exp}  &  =150\pm50\text{ MeV}%
\end{align}
We do so because the decay $K^{\ast}(1410)\rightarrow K\pi$ is well known and
the width of the rather narrow resonance $\phi(1680)$ is the sum of few decay
channels, all of them also described by our model: $\phi(1680)\rightarrow
K^{\ast}(892)K$, $\phi(1680)\rightarrow\phi(1020)\eta$, and $\phi
(1680)\rightarrow\bar{K}K.$ Moreover, $\phi(1680)\rightarrow K^{\ast}(892)K$
is reported by the PDG to be dominant: this property fits very well with our
results (details in the next subsection). Upon minimizing the function:
\begin{align}
F_{E}(g_{EPP},g_{EVP})  &  =\left(  \frac{\Gamma_{K^{\ast}(1410\rightarrow
K\pi)}-\Gamma_{K^{\ast}(1410)\rightarrow K\pi}^{\exp}}{\delta\Gamma_{K^{\ast
}(1410)\rightarrow K\pi}^{exp}}\right)  ^{2}\nonumber\\
&  +\left(  \frac{\Gamma_{\phi(1680)\rightarrow K^{\ast}(892)K}+\Gamma
_{\phi(1680)\rightarrow\phi(1020)\eta}+\Gamma_{\phi(1680)\rightarrow\bar{K}%
K}-\Gamma_{\phi(1680)}^{tot,\exp}}{\delta\Gamma_{\phi(1680)}^{tot,exp}%
}\right)  ^{2} \label{fe}%
\end{align}
we obtain:
\begin{equation}
g_{EPP}=3.66\pm0.4\text{ and }g_{EVP}=18.4\pm3.8\text{ }. \label{ge}%
\end{equation}

Similarly, in order to determine the coupling constants $g_{DPP}$ and
$g_{DVP}$ we need to choose two experimental values. For consistency, we use
the results of the experiments ASTON 84 \cite{Aston84} and ASTON 88
\cite{Aston88} in connection to the rather well known resonance $K^{\ast
}(1680)$. The values that we will use are also in agreement with the fit of
PDG, see below. The first quantity that we use is the $K\rho$ to $K\pi$ ratio
\begin{equation}
\left.  \frac{\Gamma_{K^{\ast}(1680)\rightarrow K\rho}}{\Gamma_{K^{\ast
}(1680)\rightarrow K\pi}}\right\vert _{\exp}=1.2\pm0.4\text{ by ASTON 84
\cite{Aston84}, } \label{ratiogdvp/gdpp}%
\end{equation}
which basically fixes the ratio $g_{DVP}/g_{DPP}$. Note, the fit done by the
PDG \cite{pdg} reads $0.81_{-0.09}^{+0.14}$ and is compatible with ASTON 84.
Next, we use the decay width $\Gamma_{K^{\ast}(1680)\rightarrow K\pi}$,
obtained by the two following quantities:
\begin{equation}
\left.  \frac{\Gamma_{K^{\ast}(1680)\rightarrow K\pi}}{\Gamma_{K^{\ast}%
(1680)}^{tot}}\right\vert _{\exp}=0.388\pm0.036\text{ and}\left.  \text{
}\Gamma_{K^{\ast}(1680)}^{tot}\right\vert _{\exp}=205\pm50\text{ MeV by
ASTON\ 88 \cite{Aston88}, } \label{aston88}%
\end{equation}
out of which:%
\begin{equation}
\left.  \Gamma_{K^{\ast}(1680)\rightarrow K\pi}\right\vert _{\exp}%
=79\pm21\text{ MeV from ASTON 88 \cite{Aston88}.} \label{k(1680)inkpion}%
\end{equation}
Note, the PDG quotes the fit $\Gamma_{K^{\ast}(1680)\rightarrow K\pi}%
/\Gamma_{K^{\ast}(1680)}^{tot}=0.387\pm0.026$, which is basically the value
determined in ASTON 88. The full width quoted by the PDG reads $322\pm110$ MeV
(as average) and is also compatible with ASTON 88. Finally, by using Eqs.
(\ref{ratiogdvp/gdpp}) and (\ref{k(1680)inkpion}) and performing the standard
minimization of%
\begin{align}
F_{D}(g_{DPP},g_{DVP})  &  =\left(  \frac{\frac{\Gamma_{K^{\ast}%
(1680)\rightarrow K\rho}}{\Gamma_{K^{\ast}(1680)\rightarrow K\pi}}-\left(
\frac{\Gamma_{K^{\ast}(1680)\rightarrow K\rho}}{\Gamma_{K^{\ast}%
(1680)\rightarrow K\pi}}\right)  ^{\exp}}{\delta\left(  \frac{\Gamma_{K^{\ast
}(1680)\rightarrow K\rho}}{\Gamma_{K^{\ast}(1680)\rightarrow K\pi}}\right)
}\right)  ^{2}\nonumber\\
&  +\left(  \frac{\Gamma_{K^{\ast}(1680)\rightarrow K\pi}-\Gamma_{K^{\ast
}(1680)\rightarrow K\pi}^{\exp}}{\delta\Gamma_{K^{\ast}(1680)\rightarrow K\pi
}}\right)  ^{2} \label{fd}%
\end{align}
we obtain:
\begin{equation}
g_{DPP}=7.15\pm0.94\text{ and }g_{DVP}=16.5\pm3.5\text{ .} \label{gd}%
\end{equation}
For further details on the used approach for the determination of the
parameters and for the subsequent error propagation, see Appendix C. Note, in
calculating the errors of the coupling constants of \ Eqs. (\ref{ge}) and
(\ref{gd}) we did not consider the uncertainties on the masses.

\subsection{Results for the nonet \{$\rho(1450)$, $K^{\ast}(1410)$,
$\omega(1420)$, $\phi(1680)$\}}

\begin{table}[h]
\renewcommand{\arraystretch}{1.23}
\par
\makebox[\textwidth][c] { 
\par%
\begin{tabular}
[c]{|c|c|c|}\hline
Decay process $V_{E}\rightarrow PP$ & Theory [MeV] & Experiment
[MeV]\\\hline\hline
$\rho(1450)\rightarrow\bar{K}K$ & $6.6\pm1.4$ & $<$ $6.7\pm1.0$ by\ DONANCHIE
91 \cite{Donnachie91}\\\hline
$\rho(1450)\rightarrow\pi\pi$ & $30.8\pm6.7$ & $\sim$ $27\pm4,$ seen by CLEGG
94 \cite{Clegg94}\\\hline
$K^{\ast}(1410)\rightarrow K\pi$ & $15.3\pm3.3$ & $15.3\pm3.3$ by PDG\\\hline
$K^{\ast}(1410)\rightarrow K\eta$ & $6.9\pm1.5$ & not listed in PDG\\\hline
$K^{\ast}(1410)\rightarrow K\eta^{\prime}$ & $\approx0$ & not listed in
PDG\\\hline
$\omega(1420)\rightarrow\bar{K}K$ & $5.9\pm1.3$ & not listed in PDG\\\hline
$\phi(1680)\rightarrow\bar{K}K$ & $19.8\pm4.3$ & seen by BUON 82
\cite{Buon82}\\\hline
\end{tabular}
}\caption{Decays widths of (predominantly) radially excited vector mesons into
two pseudoscalar mesons ($V_{E}\rightarrow PP$).}%
\end{table}

\begin{table}[h]
\renewcommand{\arraystretch}{1.23}
\par
\makebox[\textwidth][c] { 
\par%
\begin{tabular}
[c]{|c|c|c|}\hline
Decay process $V_{E}\rightarrow VP$ & Theory [MeV] & Experiment
[MeV]\\\hline\hline
$\rho(1450)\rightarrow\omega\pi$ & $74.7\pm31.0$ & $\sim84\pm13$ seen by CLEGG
94 \cite{Clegg94}\\\hline
$\rho(1450)\rightarrow K^{\ast}(892)K$ & $6.7\pm2.8$ & possibly seen by COAN
04 \cite{Coan04}\\\hline
$\rho(1450)\rightarrow\rho(770)\eta$ & $9.3\pm3.9$ & $<16.0\pm2.4$ by
Donnachie 91 \cite{Donnachie91}\\\hline
$\rho(1450)\rightarrow\rho(770)\eta^{\prime}$ & $\approx0$ & not listed in
PDG\\\hline
$K^{\ast}(1410)\rightarrow K\rho$ & $12.0\pm5.0$ & $<16.2\pm1.5$ by
PDG\\\hline
$K^{\ast}(1410)\rightarrow K\phi$ & $\approx0$ & not listed in PDG\\\hline
$K^{\ast}(1410)\rightarrow K\omega$ & $3.7\pm1.5$ & not listed in PDG\\\hline
$K^{\ast}(1410)\rightarrow K^{\ast}(892)\pi$ & $28.8\pm12.0$ & $>93\pm8$ by
PDG\\\hline
$K^{\ast}(1410)\rightarrow K^{\ast}(892)\eta$ & $\approx0$ & not listed in
PDG\\\hline
$K^{\ast}(1410)\rightarrow K^{\ast}(892)\eta^{\prime}$ & $\approx0$ & not
listed in PDG\\\hline
$\omega(1420)\rightarrow\rho\pi$ & $196\pm81$ & dominant, $\Gamma
_{tot}=(180-250)$ by PDG\\\hline
$\omega(1420)\rightarrow K^{\ast}(892)K$ & $2.3\pm1.0$ & not listed in
PDG\\\hline
$\omega(1420)\rightarrow\omega(782)\eta$ & $4.9\pm2.0$ & not listed in
PDG\\\hline
$\omega(1420)\rightarrow\omega(782)\eta^{\prime}$ & $\approx0$ & not listed in
PDG\\\hline
$\phi(1680)\rightarrow K\bar{K}^{\ast}$ & $110\pm46$ & dominant, $\Gamma
_{tot}=150\pm50$ by PDG\\\hline
$\phi(1680)\rightarrow\phi(1020)\eta$ & $12.2\pm5.1$ & seen by ACHASOV 14
\cite{Achasov14}\\\hline
$\phi(1680)\rightarrow\phi(1020)\eta^{\prime}$ & $\approx0$ & not listed in
PDG\\\hline
\end{tabular}
}\caption{Decays widths of (predominantly) radially excited vector mesons into
a pseudoscalar mesons and a ground-state vector meson ($V_{E}\rightarrow
VP$).}%
\end{table}

\begin{table}[h]
\renewcommand{\arraystretch}{1.23}
\par
\makebox[\textwidth][c] {%
\begin{tabular}
[c]{|c|c|c|}\hline
Decay process $V_{E}\rightarrow\gamma P$ & Theory [MeV] & Experiment
[MeV]\\\hline\hline
$\rho(1450)\rightarrow\gamma\pi$ & $0.072\pm0.042$ & not listed\\\hline
$\rho(1450)\rightarrow\gamma\eta$ & $0.23\pm0.14$ & $\sim0.2-1.5,$ see
text.\\\hline
$\rho(1450)\rightarrow\gamma\eta^{\prime}$ & $0.056\pm0.033$ & not
listed\\\hline
$K^{\ast}(1410)\rightarrow\gamma K$ & $0.18\pm0.11$ & seen, $<0.0529$ MeV PDG+
Alavi-Harati 02B \cite{Alavi-harati 02B}\\\hline
$\omega(1420)\rightarrow\gamma\pi$ & $0.60\pm0.36$ & $1.90\pm0.75,$ see
text.\\\hline
$\omega(1420)\rightarrow\gamma\eta$ & $0.023\pm0.014$ & not listed\\\hline
$\omega(1420)\rightarrow\gamma\eta^{\prime}$ & $0.0050\pm0.0030$ & not
listed\\\hline
$\phi(1680)\rightarrow\gamma\eta$ & $0.14\pm0.09$ & seen\\\hline
$\phi(1680)\rightarrow\gamma\eta^{\prime}$ & $0.076\pm0.045$ & not
listed\\\hline
\end{tabular}
}\caption{Decay widths of (predominantly) radially excited vector mesons into
a photon and a pseudoscalar meson ($V_{E}\rightarrow\gamma P$).}%
\end{table}

In Tables 6, 7, and 8 we report the results for the decays of $\{\rho
(1450),K^{\ast}(1410),\omega(1420),\phi(1680)\}$ into $PP,$ $VP$ and $\gamma
P$ pairs. An overall agreement of theory with data is visible: theoretically
large decays are clearly seen in experiments, while theoretically small decays
were generally not seen. These results show that the understanding of this
nonet as a regular $\bar{q}q$ is quite stable. Yet, there are some quantities
which deserve more detailed comments, see below. Namely, besides partial
widths, the PDG reports experimental results about ratios of different decay channels.

(i) Resonance $\rho(1450),$ strong decays. For what concerns $\rho(1450)$
various ratios can be checked. The $\pi\pi/\omega\pi$ ratio has been
determined in Ref. CLEGG 94 \cite{Clegg94}:%

\begin{equation}
\left.  \frac{\Gamma_{\rho(1450)\rightarrow\pi\pi}}{\Gamma_{\rho
(1450)\rightarrow\omega\pi}}\right\vert _{\exp}\sim0.32\text{ by CLEGG 94
\cite{Clegg94}.}%
\end{equation}
The corresponding theoretical $0.41\pm0.20$ agrees well with CLEGG 94. Note,
this ratio depends on the ratio of coupling constant $g_{EPP}/g_{EVP}$ and
therefore represents an independent confirmation of this quantity. Along the
same line, the ratio%

\begin{equation}
\left.  \frac{\Gamma_{\rho(1450)\rightarrow\pi\pi}}{\Gamma_{\rho
(1450)\rightarrow\eta\rho}}\right\vert _{\exp}=1.3\pm0.4\text{ AULCHENKO 15
\cite{Aulchenko15}}%
\end{equation}
is in qualitative agreement with the theoretical value of $3.3\pm1.6.$ Going
on, the upper limit for the ratio%
\begin{equation}
\left.  \frac{\Gamma_{\rho(1450)\rightarrow KK}}{\Gamma_{\rho(1450)\rightarrow
\omega\pi}}\right\vert _{\exp}<0.08\text{ DONNACHIE 91 \cite{Donnachie91}}%
\end{equation}
is also compatible with the theoretical value of $0.088\pm0.043.$ Summarizing,
$g_{EPP}/g_{EVP}\simeq1/5$ is in good agreement with various experimental
results of $\rho(1450)$. Next, we consider%

\begin{equation}
\left.  \frac{\Gamma_{\rho(1450)\rightarrow\eta\rho}}{\Gamma_{\rho
(1450)\rightarrow\omega\pi}}\right\vert _{\exp}=\left\{
\begin{tabular}
[c]{l}%
$0.081\pm0.020\text{ AULCHENKO 15 \cite{Aulchenko15}}$\\
$\sim0.21$ DONNACHIE 91 \cite{Donnachie91}\\
$>2$ FUKUI 91 \cite{Fukui91}%
\end{tabular}
\ \ \ \ \right.
\end{equation}
which shall be compared to our theoretical result of $\approx0.12.$ This value
is in good agreement with the latest determination of Ref. AULCHENKO 15
\cite{Aulchenko15} and also with DONNACHIE 91 \cite{Donnachie91}. On the
contrary, the lower limit of FUKUI 91 \cite{Fukui91} is in disagreement with
the other experiments as well as with theory. Here, the theoretical result is
parameter independent: this ratio is purely fixed by flavour symmetry and
phase space. Due to the fact that in Sec. 3.1 we used only the errors of the
decay width to determine the parameter's errors, no theoretical error for this
ratio can be determined. Of course, the uncertainties of other quantities
(such as the masses) and the validity of the employed Lagrangians (see the
discussion in\ Sec. 2.2) induce an error for this quantity (with the estimate
of about $10$-$20\%$, which is the expected precision of our effective model).
The very same comment will apply to all the ratios of the type $PP/PP$ and
$PV/PV$, which are independent on the coupling constants of our approach.

As a concluding remark, the interpretation of $\rho(1450)$ as an excited
$\rho$ meson is in well agreement with the present data of various
experiments. In this respect, a lighter resonance $\rho(1290),$ see e.g. Refs.
\cite{coito1,coito2,nazari}, is not needed in the $\bar{q}q$ assignment.
Eventually, such a lighter resonance can emerge as a companion pole
\cite{pennington,wolkanowski,milenathomas} once loops are included.

(ii) Resonance $K^{\ast}(1410),$ strong decays. The resonance $K^{\ast}(1410)$
is well established, both experimentally and on the lattice \cite{prelovsek}.
Its decay into $K\pi$ reported in Table 6 turns out to be correct because this
branching ratio was used to fix the parameters (see Sec. 3.1). However, the
decay $K^{\ast}(1410)\rightarrow K^{\ast}(892)\pi$ is too small when compared
to the summary of the PDG: roughly, an overestimation of a factor three is
present. We now discuss the ratios of $K^{\ast}(1410)$.\ We start with%
\begin{equation}
\left.  \frac{\Gamma_{K^{\ast}(1410)\rightarrow\rho K}}{\Gamma_{K^{\ast
}(1410)\rightarrow K^{\ast}(892)\pi}}\right\vert _{\exp}<0.17\text{ ASTON 84
\cite{Aston84},}%
\end{equation}
which should be compared to $\approx0.42$, which is too large. This ratio is
fixed by flavour symmetry and is independent on the parameters, hence a
disagreement is quite surprising and should be clarified in the future. Next,
we consider%

\begin{equation}
\left.  \frac{\Gamma_{K^{\ast}(1410)\rightarrow\pi K}}{\Gamma_{K^{\ast
}(1410)\rightarrow K^{\ast}(892)\pi}}\right\vert _{\exp}<0.16\text{ ASTON 84
\cite{Aston84},}%
\end{equation}
that should be compared with $0.53\pm0.26.$ Indeed, this is the seed of the
disagreement concerning the decay $K^{\ast}(1410)\rightarrow K^{\ast}(892)\pi$
of Table 7. Namely, the value by ASTON 84 has been used to set the upper limit
quoted in the PDG.

(iii) Resonance $\omega(1420),$ strong decays. The decays of the resonance
$\omega(1420)$ are in agreement with data. The decay $\omega(1420)\rightarrow
K^{\ast}(892)K$ is theoretically the largest one (in agreement with the PDG,
which classifies this decay as dominant). All the other decay rates are quite
small (few MeV) and were not yet discovered experimentally (hence, our results
are predictions). From the 2017 update of the PDG one can use the values
\begin{align}
\left.  \frac{\Gamma_{\omega(1420)\rightarrow\omega\eta}}{\Gamma
_{\omega(1420)}^{tot}}\frac{\Gamma_{\omega(1420)\rightarrow e^{+}e^{-}}%
}{\Gamma_{\omega(1420)}^{total}}\right\vert _{\exp}  &  =(1.6_{-0.07}%
^{+0.09})\cdot10^{-8}\text{ by ACHASOV 16B \cite{Achasov16B}}%
\label{omegarhopion}\\
\left.  \frac{\Gamma_{\omega(1420)\rightarrow\rho\pi}}{\Gamma_{\omega
(1420)}^{tot}}\frac{\Gamma_{\omega(1420)\rightarrow e^{+}e^{-}}}%
{\Gamma_{\omega(1420)}^{total}}\right\vert _{\exp}  &  =\left(  0.73\pm
0.08\right)  \cdot10^{-6}\text{ by AULCHENKO 15A \cite{Aulchenko15A}}
\label{omegarhopionepem}%
\end{align}
to extract (in the numerator the average of errors is used)
\begin{equation}
\left.  \frac{\Gamma_{\omega(1420)\rightarrow\omega\eta}}{\Gamma
_{\omega(1420)\rightarrow\rho\pi}}\right\vert _{\exp}=\frac{(1.6\pm
0.08)\cdot10^{-8}}{\left(  0.73\pm0.08\right)  \cdot10^{-6}}=0.021\pm0.001
\end{equation}
This result compares very well to our theoretical result $\approx0.025$.
(Note, for the quantity in Eq. (\ref{omegarhopion}) other values are listed in
the PDG: they are all compatible to Eq. (\ref{omegarhopion}) and to each other.)

(iv) Resonance $\phi(1680),$ strong decays. The partial decay widths of the
resonance $\phi(1680)$ fits reported in Table 6 and 7 show that the channel
$\phi(1680)\rightarrow KK^{\ast}(892)$ is dominant, in agreement with the PDG
quote. Moreover, the following ratios can be experimentally obtained and both
of them are compatible with theory:
\begin{equation}
\left.  \frac{\Gamma_{\phi(1680)\rightarrow K\bar{K}}}{\Gamma_{\phi
(1680)\rightarrow K^{\ast}(892)K}}\right\vert _{\exp}=0.07\pm0.01\text{ BUON
82 \cite{Buon82}}%
\end{equation}
is in good agreement with the theoretical value (dependent on the ratio of
couplings $g_{EPP}/g_{EVP}$) of $0.18\pm0.09$; similarly, the parameter
independent ratio%

\begin{equation}
\left.  \frac{\Gamma_{\phi(1680)\rightarrow\eta\phi}}{\Gamma_{\phi
(1680)\rightarrow K^{\ast}(892)K}}\right\vert _{\exp}=0.07\pm0.01\text{ AUBERT
08S \cite{Aubert08S}}%
\end{equation}
agrees with the theoretical value of $\approx0.11.$

(iv) Radiative decays. As a last step, we discuss also the decay rates of
excited vector mesons into a photon and pseudoscalar mesons. These radiative
decays were determined by using `Vector Meson Dominance' without the need of
any new parameter. The radiative decays of \{$\rho(1450)$, $K^{\ast}(1410)$,
$\omega(1420)$, $\phi(1680)$\} are still poorly determined experimentally, but
the theoretically predicted sizable decays were seen in experiments. In two
cases, numerical values can be extracted (in other cases, our theoretical
results represent predictions). The decay $\rho(1450)\rightarrow\gamma\eta$
can be estimated by using
\begin{equation}
\left.  \Gamma_{\rho(1450)\rightarrow\gamma\eta}\frac{\Gamma_{\rho
(1450)\rightarrow e^{+}e^{-}}}{\Gamma_{\rho(1450)}^{total}}\right\vert _{\exp
}=2.2\pm0.5\pm0.3\text{ eV by AKHMETSHIN 01B \cite{Akhmetshin01B},}%
\end{equation}
and
\begin{equation}
\left.  \Gamma_{\rho(1450)\rightarrow\pi\pi}\frac{\Gamma_{\rho
(1450)\rightarrow e^{+}e^{-}}}{\Gamma_{\rho(1450)}^{total}}\right\vert _{\exp
}=\left\{
\begin{tabular}
[c]{l}%
$\ 0.12$ keV by DIEKMANN 88 \cite{Diekmann88}\\
$0.027_{-0.010}^{+0.015}$ keV by KURDADZE 83 \cite{Kurdadze83}%
\end{tabular}
\ \ \ \ \ \ \right.  .
\end{equation}
as well as $\Gamma_{\rho(1450)\rightarrow\pi\pi}\approx84$ MeV (see CLEGG 94
\cite{Clegg94}). One obtains:
\begin{equation}
\left.  \Gamma_{\rho(1450)\rightarrow\gamma\eta}\right\vert _{\exp}%
\approx\left\{
\begin{tabular}
[c]{l}%
$1.5$ MeV\\
$0.2$ MeV
\end{tabular}
\ \ \ \ \ \ \right.  \text{ ,}%
\end{equation}
which is reported in Table 8. The error unfortunately cannot be determined
since CLEGG 94 did not report any and there are no other determinations of
$\Gamma_{\rho(1450)\rightarrow\pi\pi}$. Nevertheless, the qualitative
agreement of the second determination with the theoretical result
($0.23\pm0.14$ MeV) is rather promising. Similarly, for what concerns
$\omega(1420)\rightarrow\gamma\pi^{0}$ we use
\begin{equation}
\left.  \frac{\Gamma_{\omega(1420)\rightarrow\gamma\pi^{0}}}{\Gamma
_{\omega(1420)}^{total}}\frac{\Gamma_{\omega(1420)\rightarrow e^{+}e^{-}}%
}{\Gamma_{\omega(1420)}^{total}}\right\vert _{\exp}=2.03_{-0.75}^{+0.70}%
\cdot10^{-8}\text{ by AKHMETSHIN 05 \cite{Akhmetshin05}}%
\end{equation}
together with
\begin{equation}
\left.  \frac{\Gamma_{\omega(1420)\rightarrow e^{+}e^{-}}}{\Gamma
_{\omega(1420)}^{total}}\right\vert _{\exp}=\left(  23\pm1\right)
\cdot10^{-7}\text{ by HENNER 02 \cite{Henner02}} \label{omegaepem}%
\end{equation}
to obtain:
\begin{equation}
\left.  \Gamma_{\omega(1420)\rightarrow\gamma\pi^{0}}\right\vert _{\exp
}=\left(  1.9\pm0.75\right)  \text{ MeV.}%
\end{equation}
(Note, the ratio (\ref{omegaepem}) was also estimated by ACHASOV\ 03D
\cite{Achasov03D} to be $6.6$, but without errors, therefore we chose to use
HENNER 02. Moreover, the value $6.6$ combined with Eq. (\ref{omegarhopionepem}%
) would deliver a ratio $\Gamma_{\omega(1420)\rightarrow\rho\pi}%
/\Gamma_{\omega(1420)}^{tot}$ larger than one, a result which is not
consistent.) Also in this case, the agreement with the theoretical result
$\Gamma_{\omega(1420)\rightarrow\gamma\pi^{0}}=0.60\pm0.36$ MeV is
interesting. Quite remarkably, the decay $\omega(1420)\rightarrow\gamma\pi
^{0}$ is the strongest radiative decay that the theory predicts and
correspondingly there is a sizable value that can be extracted by the present
experimental information.

Finally, the upper limit of decay $K^{\ast}(1410)\rightarrow K\gamma$ as
quoted by the PDG (determined in the work of Ref. ALAVI-HARATI 02B
\cite{Alavi-harati 02B}) is $0.052$ MeV; this value is compatible with the
present theoretical result. In future experiments it should be possible to
determine this quantity. The decay width $\phi(1680)\rightarrow\gamma\eta$ has
been also seen experimentally by ACHASOV 14 \cite{Achasov14}, but no value is reported.

Summarizing, the overall agreement of theory with data is stable and confirms
that the first nonet of excited vectors \{$\rho(1450)$, $K^{\ast}(1410)$,
$\omega(1420)$, $\phi(1680)$\} is a standard $\bar{q}q$ nonet, predominantly
corresponding to the first radial excitation of vector mesons. Future
experimental results are expected to come and further checks will be possible
in the near (GlueX and CLAS12) and less near (PANDA) future.

\subsection{Results for the nonet \{$\rho(1700)$, $K^{\ast}(1680)$,
$\omega(1650)$, $\phi(???)\equiv\phi(1930)$\}}

\begin{table}[h]
\renewcommand{\arraystretch}{1.23}
\par
\makebox[\textwidth][c] {%
\begin{tabular}
[c]{|c|c|c|}\hline
Decay process $V_{D}\rightarrow PP$ & Theory [MeV] & Experiment
[MeV]\\\hline\hline
$\rho(1700)\rightarrow\bar{K}K$ & $40\pm11$ & $8.3_{-8.3}^{+10}$ MeV, see
text.\\\hline
$\rho(1700)\rightarrow\pi\pi$ & $140\pm37$ & $75\pm30$ by BECKER 79
\cite{Becker79}\\\hline
$K^{\ast}(1680)\rightarrow K\pi$ & $82\pm22$ & $125\pm43$ by PDG\\\hline
$K^{\ast}(1680)\rightarrow K\eta$ & $52\pm14$ & not listed in PDG\\\hline
$K^{\ast}(1680)\rightarrow K\eta^{\prime}$ & $0.72\pm0.02$ & not listed in
PDG\\\hline
$\omega(1650)\rightarrow\bar{K}K$ & $37\pm10$ & not listed in PDG\\\hline
$\phi(1930)\rightarrow\bar{K}K$ & $104\pm28$ & resonance not yet known\\\hline
\end{tabular}
}\caption{Decays widths of (predominantly) orbitally excited vector mesons
into two pseudoscalar mesons ($V_{D}\rightarrow PP$).}%
\end{table}

\begin{table}[h]
\renewcommand{\arraystretch}{1.23}
\par
\makebox[\textwidth][c] {
\begin{tabular}
[c]{|c|c|c|}\hline
Decay process $V_{D}\rightarrow VP$ & Theory [MeV] & Experiment
[MeV]\\\hline\hline
$\rho(1700)\rightarrow\omega\pi$ & $140\pm59$ & seen, see text\\\hline
$\rho(1700)\rightarrow K^{\ast}(892)K$ & $56\pm23$ & $83\pm66$ MeV, see
text.\\\hline
$\rho(1700)\rightarrow\rho\eta$ & $41\pm17$ & $68\pm42$ MeV,see text\\\hline
$\rho(1700)\rightarrow\rho\eta^{\prime}$ & $\approx0$ & not listed in
PDG\\\hline
$K^{\ast}(1680)\rightarrow K\rho$ & $64\pm27$ & $101\pm35$ by PDG\\\hline
$K^{\ast}(1680)\rightarrow K\phi$ & $13\pm6$ & not listed in PDG\\\hline
$K^{\ast}(1680)\rightarrow K\omega$ & $21\pm9$ & not listed in PDG\\\hline
$K^{\ast}(1680)\rightarrow K^{\ast}(892)\pi$ & $81\pm34$ & $96\pm33$ by
PDG\\\hline
$K^{\ast}(1680)\rightarrow K^{\ast}(892)\eta$ & $0.5\pm0.2$ & not listed in
PDG\\\hline
$K^{\ast}(1680)\rightarrow K^{\ast}(892)\eta^{\prime}$ & $\approx0$ & not
listed in PDG\\\hline
$\omega(1650)\rightarrow\rho\pi$ & $370\pm156$ & $\sim205,$ $154\pm44,$
$\sim273,$ $120\pm18$, see text\\\hline
$\omega(1650)\rightarrow K^{\ast}(892)K$ & $42\pm18$ & not listed in
PDG\\\hline
$\omega(1650)\rightarrow\omega(782)\eta$ & $32\pm13$ & $\sim100,$ $56\pm30,$
see text.\\\hline
$\omega(1650)\rightarrow\omega(782)\eta^{\prime}$ & $\approx0$ & not listed in
PDG\\\hline
$\phi(1930)\rightarrow K\bar{K}^{\ast}$ & $260\pm109$ & resonance not yet
known\\\hline
$\phi(1930)\rightarrow\phi(1020)\eta$ & $67\pm28$ & resonance not yet
known\\\hline
$\phi(1930)\rightarrow\phi(1020)\eta^{\prime}$ & $\approx0$ & resonance not
yet known\\\hline
\end{tabular}
}\caption{Decays widths of (predominantly) orbitally excited vector mesons
into a pseudoscalar mesons and a ground-state vector meson ($V_{D}\rightarrow
VP$).}%
\end{table}

\begin{table}[h]
\renewcommand{\arraystretch}{1.23}
\par
\makebox[\textwidth][c] {
\begin{tabular}
[c]{|c|c|c|}\hline
Decay process $V_{D}\rightarrow\gamma P$ & Theory [MeV] & Experiment
[MeV]\\\hline\hline
$\rho(1700)\rightarrow\gamma\pi$ & $0.095\pm0.058$ & not listed\\\hline
$\rho(1700)\rightarrow\gamma\eta$ & $0.35\pm0.21$ & not listed\\\hline
$\rho(1700)\rightarrow\gamma\eta^{\prime}$ & $0.13\pm0.08$ & not
listed\\\hline
$K^{\ast}(1680)\rightarrow\gamma K$ & $0.30\pm0.18$ & not listed\\\hline
$\omega(1650)\rightarrow\gamma\pi$ & $0.78\pm0.47$ & not listed\\\hline
$\omega(1650)\rightarrow\gamma\eta$ & $0.035\pm0.021$ & not listed\\\hline
$\omega(1650)\rightarrow\gamma\eta^{\prime}$ & $0.012\pm0.007$ & not
listed\\\hline
$\phi(1930)\rightarrow\gamma\eta$ & $0.19\pm0.12$ & resonance not yet
known\\\hline
$\phi(1930)\rightarrow\gamma\eta^{\prime}$ & $0.13\pm0.08$ & resonance not yet
known\\\hline
\end{tabular}
}\caption{Decay widths of (predominantly) orbitally excited vector mesons into
a photon and a pseudoscalar meson ($V_{D}\rightarrow\gamma P$).}%
\end{table}

In Tables 9, 10, 11 we report the results for the second nonet of excited
vector mesons \{$\rho(1700)$, $K^{\ast}(1680)$, $\omega(1650)$, $\phi
(???)\equiv\phi(1930)$\}. With some exceptions to be discussed later on, there
is also in this case an overall qualitative agreement of theory with data. One
may therefore conclude that the assignment of these mesons to a nonet of
orbitally excited vector mesons is viable. Next, we concentrate on the
detailed description of the results and to the comparison of numerous ratios
listed in the PDG.

(i) Resonance $\rho(1700),$ strong decays. The total width of the resonance
$\rho(1700)$ as resulting from our theoretical analysis reads $417\pm147$
which is in agreement with the PDG estimate of $250\pm100$ MeV. The
theoretical results show a slight overestimation of $PP$ decays. Namely, while
the $\pi\pi$ channel is in agreement with BECKER 79 \cite{Becker79} reported
in Table 9, there are additional measurements of this channel in older
experiments:
\begin{equation}
\left.  \Gamma_{\rho(1700)\rightarrow\pi\pi}\right\vert _{\exp}=\left\{
\begin{tabular}
[c]{l}%
$56\pm29$ MARTIN 78C \cite{Martin78C}\\
$75\pm32$ FROGGATT 77 \cite{Froggatt77}\\
$63\pm30$ HYAMS\ 73 \cite{Hyams73}%
\end{tabular}
\ \ \ \ \ \ \ \right.  \text{ ;}%
\end{equation}
overall it looks compatible, but a new experimental determination would be useful.

For the $KK$ channel, we combine
\begin{equation}
\left.  \frac{\Gamma_{\rho(1700)\rightarrow KK}}{\Gamma_{\rho(1700)\rightarrow
2(\pi^{+}\pi^{-})}}\right\vert _{\exp}=0.015\pm0.010\text{ DELCOURT\ 81B
\cite{Delcourt81B},}%
\end{equation}

\begin{equation}
\left.  \frac{\Gamma_{\rho(1700)\rightarrow\pi\pi}}{\Gamma_{\rho
(1700)\rightarrow2(\pi^{+}\pi^{-})}}\right\vert _{\exp}=0.13\pm0.05\text{
ASTON 80 \cite{Aston80},}%
\end{equation}
and
\begin{equation}
\left.  \frac{\Gamma_{\rho(1700)\rightarrow\pi\pi}}{\Gamma_{\rho(1700)}^{tot}%
}\right\vert _{\exp}=0.287_{-0.042}^{+0.043}\pm0.05\text{ BECKER 79
\cite{Becker79},}%
\end{equation}
in order to obtain:%
\begin{equation}
\left.  \Gamma_{\rho(1700)\rightarrow KK}\right\vert _{\exp}=8.3_{-8.3}%
^{+10.4}\text{ MeV. }%
\end{equation}
This is the result reported in Table 9. It is smaller than the theoretical
result, but large errors are present.

In a similar way, we use%

\begin{equation}
\left.  \frac{\Gamma_{\rho(1700)\rightarrow KK^{\ast}(892)}}{\Gamma
_{\rho(1700)\rightarrow2(\pi^{+}\pi^{-})}}\right\vert _{\exp}=0.15\pm
0.03\text{ DELCOURT\ 81B \cite{Delcourt81B},}%
\end{equation}
to obtain%
\begin{equation}
\left.  \Gamma_{\rho(1700)\rightarrow KK^{\ast}(892)}\right\vert _{\exp}%
=83\pm66\text{ MeV }%
\end{equation}
reported in Table 10. Although the resulting experimental error is very large,
the result is compatible with the theoretical value of $56\pm23$ MeV. Note,
the decay mode $\rho(1700)\rightarrow KK^{\ast}(892)$ was also possibly seen
by COAN 04 \cite{Coan04} and clearly seen in radiative decays by BIZOT 80
\cite{Bizot80} and DELCOURT 81B \cite{Delcourt81B}, but one cannot use those
results to obtain an independent determination of this partial width.

The last decay that can be determined along the same procedure is
$\rho(1700)\rightarrow\rho\eta.$ Upon using%

\begin{equation}
\left.  \frac{\Gamma_{\rho(1700)\rightarrow\rho\eta}}{\Gamma_{\rho
(1700)\rightarrow2(\pi^{+}\pi^{-})}}\right\vert _{\exp}=0.123\pm0.027\text{
DELCOURT\ 82 \cite{Delcourt82},}%
\end{equation}
we derive
\begin{equation}
\left.  \Gamma_{\rho(1700)\rightarrow\rho\eta}\right\vert _{\exp}%
=68\pm42\text{ MeV,} \label{rhoeta}%
\end{equation}
reported in Table 10. Again, the error is large, but the value fits well with
theory. [For completeness, it should be also stressed that $\Gamma
_{\rho(1700)\rightarrow\rho\eta}/\Gamma_{\rho(1700)}^{tot}$ was determined to
be $<0.04$ by DONNACHIE 87B \cite{Donnachie87B}, out of which $\Gamma
_{\rho(1700)\rightarrow\rho\eta}<10\pm4$. This result is not compatible with
theory and with Eq. (\ref{rhoeta}).]

We now turn to the discussion of ratios. To this end, we use the reported
decay widths involving the dilepton pair $e^{+}e^{-}$, $\Gamma_{\rho
(1700)\rightarrow MM}\cdot\frac{\Gamma_{\rho(1700)\rightarrow e^{+}e^{-}}%
}{\Gamma_{\rho(1700)}^{tot}}$ , and decay ratios involving $2(\pi^{+}\pi^{-}%
)$, $\frac{\Gamma_{\rho(1700)\rightarrow MM}}{\Gamma_{\rho(1700)\rightarrow
2(\pi^{+}\pi^{-})}}$, where $MM$ is a certain meson-meson channel $(PP$ or
$VP).$ We first study $PP/PP$ ratios, then $PP/PV,$ and finally $PV/PV.$

For what concerns the (parameter independent) $\pi\pi/KK$ ratio, we have
(first two ratios from $e^{+}e^{-}$, third one from $2(\pi^{+}\pi^{-})$):%

\begin{equation}
\frac{\Gamma_{\rho(1700)\rightarrow\pi\pi}}{\Gamma_{\rho(1700)\rightarrow KK}%
}=\left\{
\begin{tabular}
[c]{l}%
$\sim3.7$ DIEKMAN 88 \cite{Diekmann88} + BIZOT 80 \cite{Bizot80}\\
$0.83\pm0.82$ KURDADZE 83 \cite{Kurdadze83}+ BIZOT 80 \cite{Bizot80}\\
$8.7\pm6.7$ ASTON 80 \cite{Aston80} + DELCOURT 81B \cite{Delcourt81B}%
\end{tabular}
\ \ \ \ \ \ \right.  \text{ ,}%
\end{equation}
that shall be compared to the theoretical result of $\approx3.5$, which is in
rough agreement with experiment (especially with the first determination above).

For what concerns $\pi\pi/\pi\rho,$ one obtains (first two ratios from
$e^{+}e^{-}$, third one from $2(\pi^{+}\pi^{-})$):%

\begin{equation}
\left.  \frac{\Gamma_{\rho(1700)\rightarrow\pi\pi}}{\Gamma_{\rho
(1700)\rightarrow\eta\rho}}\right\vert _{\exp}=\left\{
\begin{tabular}
[c]{l}%
$\sim18$ DIEKMANN 88 \cite{Diekmann88} + ANTONELLI 88 \cite{Antonelli88}\\
$4.1\pm2.7$ KURDADZE 83 \cite{Kurdadze83} + ANTONELLI\ 88 \cite{Antonelli88}\\
$1.1\pm0.47$ ASTON\ 80 \cite{Aston80} + DELCOURT 82 \cite{Delcourt82}%
\end{tabular}
\ \ \ \ \ \ \right.  \text{ .}%
\end{equation}
The theoretical result $3.4\pm1.1$ fits quite well with the second entry.
There is also agreement with the last experimental result quoted above. On the
contrary, the first entry (without error) is not compatible with theory and
with the other experimental values.

Next, for the $KK/\eta\rho$ ratio (first result from $e^{+}e^{-},$ second from
$2(\pi^{+}\pi^{-})$) we get:%

\begin{equation}
\left.  \frac{\Gamma_{\rho(1700)\rightarrow KK}}{\Gamma_{\rho(1700)\rightarrow
\eta\rho}}\right\vert _{\exp}=\left\{
\begin{tabular}
[c]{l}%
$5.0\pm4.7$ BIZOT 80 \cite{Bizot80}+ ANTONELLI 88 \cite{Antonelli88}\\
$0.12\pm0.09$ DELCOURT 81B \cite{Delcourt81B} + ANTONELLI\ 88
\cite{Antonelli88}%
\end{tabular}
\ \ \ \ \ \ \right.
\end{equation}
The corresponding theoretical width of $0.98\pm0.33$ is compatible with the
first entry (due to the larger experimental errors) but not with the second.
Also in this case, the experimental values do not agree with each other.

Going further, we discuss the ratio $KK/K^{\ast}(892)K,$ which is quite
problematic (first result from $e^{+}e^{-},$ second from $2(\pi^{+}\pi^{-})$,
third one listed in the PDG):%

\begin{equation}
\left.  \frac{\Gamma_{\rho(1700)\rightarrow KK}}{\Gamma_{\rho(1700)\rightarrow
K^{\ast}(892)K}}\right\vert _{\exp}=\left\{
\begin{tabular}
[c]{l}%
$0.11\pm0.10$ BIZOT 80 \cite{Bizot80}\\
$0.10\pm0.07$ DELCOURT 81 B \cite{Delcourt81B}\\
$0.052\pm0.026$ BUON 82 \cite{Buon82}%
\end{tabular}
\ \ \ \ \ \ \right.  . \label{kkk*k}%
\end{equation}
The corresponding theoretical result reads $0.71\pm0.24$. Hence, we have a
mismatch of the listed experimental ratios with our value. The reason for this
mismatch is easy to understand: ratios of the type $PP/VP$ depend solely on
the ratio of coupling constants $g_{DPP}/g_{DVP},$ which is fixed by Eq.
(\ref{ratiogdvp/gdpp}) in which the results of ASTON 84 \cite{Aston84} and
ASTON 88 \cite{Aston88} are used (see Sec. 3.1). There is no way to bring
those experimental result and the ones of Eq. (\ref{kkk*k}) in agreement with
each other. Since the ratio in Eq. (\ref{ratiogdvp/gdpp}) seems to be based on
a quite solid result, we tend to believe that a new determination of
$KK/K^{\ast}(892)K$ is necessary.

Finally, we turn to the (parameter independent) $K^{\ast}(892)K/\eta\rho$
ratio (first result from $e^{+}e^{-}$, second from $2(\pi^{+}\pi^{-})$):%

\begin{equation}
\left.  \frac{\Gamma_{\rho(1700)\rightarrow K^{\ast}(892)K}}{\Gamma
_{\rho(1700)\rightarrow\eta\rho}}\right\vert _{\exp}=\left\{
\begin{tabular}
[c]{l}%
$43\pm21$ BIZOT 80 \cite{Bizot80} + ANTONELLI 88 \cite{Antonelli88}\\
$1.22\pm0.27$ DELCOURT 81B \cite{Delcourt81B} + DELCOURT 82 \cite{Delcourt82}%
\end{tabular}
\ \ \ \ \ \ \ \ \right.
\end{equation}
Unfortunately, the two values are not consistent with each other. The
theoretical result reads $\approx1.37$ fits quite well with the second entry.
This result shows that $\Gamma_{\rho(1700)\rightarrow K^{\ast}(892)K}$ is
possibly overestimated by the quantity $\Gamma_{\rho(1700)\rightarrow K^{\ast
}(892)K}\cdot\frac{\Gamma_{\rho(1700)\rightarrow e^{+}e^{-}}}{\Gamma
_{\rho(1700)}^{tot}}=0.305\pm0.071$ reported by BIZOT 80 \cite{Bizot80}. A
smaller value of the latter would lead to a better agreement with our
theoretical results. This comment also applies for the disagreement with the
ratio reported in Eq. (\ref{kkk*k}).

(ii) Resonance $K^{\ast}(1680),$ strong decays. We now discuss the resonance
$K^{\ast}(1680),$ which is experimentally rather well known. The decay widths
fit well with the experiment (in agreement with the fact that we used one of
them to fix the strength of the parameters, see Sec. 3.1). In addition, we can
study two ratios.

The $K\pi/K^{\ast}(892)\pi$ ratio is determined by the PDG and ASTON 84 (both
entries bold) as%

\begin{equation}
\left.  \frac{\Gamma_{K^{\ast}(1680)\rightarrow K\pi}}{\Gamma_{K^{\ast
}(1680)\rightarrow K^{\ast}(892)\pi}}\right\vert _{\exp}=\left\{
\begin{tabular}
[c]{l}%
$1.30_{-0.14}^{+0.23}$ fit by PDG\\
$2.8\pm1.1$ by ASTON\ 84 \cite{Aston84}%
\end{tabular}
\ \ \ \ \ \right.  .
\end{equation}
The two values do not agree well with each other, but due to large errors they
are not incompatible. The theoretical result $1.01\pm0.34$ fits well with the
PDG value.

The second ratio, $K\rho/K^{\ast}(892)\pi,$ reads%

\begin{equation}
\left.  \frac{\Gamma_{K^{\ast}(1680)\rightarrow K\rho}}{\Gamma_{K^{\ast
}(1680)\rightarrow K^{\ast}(892)\pi}}\right\vert _{\exp}=\left\{
\begin{tabular}
[c]{l}%
$1.05_{-0.11}^{+0.27}$fit by PDG\\
$0.97\pm0.09_{-0.10}^{+0.30}$ by ASTON\ 87 \cite{Aston87}%
\end{tabular}
\ \ \ \ \ \ \right.  .
\end{equation}
The theoretical value is $\approx0.79$ fits well with both entries.

(iii) Resonance $\omega(1650),$ strong decays. The dominant decay of
$\omega(1650)$ is represented by the mode $\omega(1650)\rightarrow\rho\pi.$
The qualitative picture agrees well with the theory (see table 10). Yet, the
theoretical result has a very large error. The decay width $\Gamma
_{\omega(1650)\rightarrow\rho\pi}$ can be determined by using the ratios
\begin{equation}
\left.  \frac{\Gamma_{\omega(1650)\rightarrow\rho\pi}}{\Gamma_{\omega
(1650)}^{tot}}\right\vert _{\exp}=\left\{
\begin{array}
[c]{c}%
\sim0.65\text{ ACHASOV 03D \cite{Achasov03D}}\\
0.380\pm0.014\text{ HENNER 02 \cite{Henner02}}%
\end{array}
\right.
\end{equation}
together with $\Gamma_{\omega(1650)}^{tot}=315\pm35$ MeV \cite{pdg}, finding:%
\begin{equation}
\left.  \Gamma_{\omega(1650)\rightarrow\rho\pi}\right\vert _{\exp}=\left\{
\begin{array}
[c]{c}%
\sim205\text{ MeV ACHASOV 03D \cite{Achasov03D}}\\
120\pm18\text{ MeV HENNER 02 \cite{Henner02}.}%
\end{array}
\right.  \text{ .} \label{rhopion1}%
\end{equation}
The theoretical result is in agreement with the upper determination but
overestimate the latter. (The latter value would point to a smaller value of
the parameter $g_{DVP}$)$.$ Yet, other determinations can be obtained by using%
\begin{equation}
\frac{\Gamma_{\omega(1650)\rightarrow\rho\pi}}{\Gamma_{\omega(1650)}^{tot}%
}\frac{\Gamma_{\omega(1650)\rightarrow e^{+}e^{-}}}{\Gamma_{\omega
(1650)}^{tot}}=1.56\pm0.23\text{ AULCHENKO 15A \cite{Aulchenko15A}}
\label{rad16501}%
\end{equation}
together with
\begin{equation}
\left.  \frac{\Gamma_{\omega(1650)\rightarrow e^{+}e^{-}}}{\Gamma
_{\omega(1650)}^{tot}}\right\vert _{\exp}=\left\{
\begin{array}
[c]{c}%
\sim18\text{ ACHASOV 03D \cite{Achasov03D}}\\
32\pm1\text{ HENNER 02 \cite{Henner02}}%
\end{array}
\right.
\end{equation}
out of which:%
\begin{equation}
\Gamma_{\omega(1650)\rightarrow\rho\pi}=\left\{
\begin{array}
[c]{c}%
\sim273\text{ MeV, ACHASOV 03D \cite{Achasov03D} + AULCHENKO 15A
\cite{Aulchenko15A}}\\
154\pm44\text{ HENNER 02 \cite{Henner02} + AULCHENKO 15A \cite{Aulchenko15A}}%
\end{array}
\right.  \text{ .}%
\end{equation}
The results still have large error and a clear outcome is difficult to assess.
Surely, this decay width is large and is the dominant decay channel of
$\omega(1650).$

Following the same procedure, by using
\begin{equation}
\left.  \frac{\Gamma_{\omega(1650)\rightarrow\omega\eta}}{\Gamma
_{\omega(1650)}^{tot}}\frac{\Gamma_{\omega(1650)\rightarrow e^{+}e^{-}}%
}{\Gamma_{\omega(1650)}^{tot}}\right\vert _{\exp}=0.57\pm0.06\text{ AUBERT 06D
\cite{Aubert06D}, } \label{rad16502}%
\end{equation}
we obtain:
\begin{equation}
\left.  \Gamma_{\omega(1650)\rightarrow\omega\eta}\right\vert _{\exp}=\left\{
\begin{array}
[c]{c}%
\sim100\text{ MeV, ACHASOV 03D \cite{Achasov03D} +AUBERT 06D \cite{Aubert06D}
}\\
56\pm30\text{ MeV, HENNER 02 \cite{Henner02} + AUBERT 06D \cite{Aubert06D}}%
\end{array}
\right.  \text{ ,}%
\end{equation}
which shall be compared with the theoretical result of $32\pm13.$ Hence, it
fits quite well with the second.

Finally, we can use Eqs. (\ref{rad16501}) and (\ref{rad16502}) to determine
the ratio%

\begin{equation}
\left.  \frac{\Gamma_{\omega(1650)\rightarrow\omega\eta}}{\Gamma
_{\omega(1650)\rightarrow\rho\pi}}\right\vert _{\exp}=0.365\pm0.054\text{ }%
\end{equation}
which is somewhat larger than the theoretical value $\approx0.086$.

(iv) Putative resonance $\phi(1930),$ strong decays. This resonance has not
been found yet. The results of Table 9 and 10 are therefore predictions. For
the reader's convenience we summarize them in Table 12. Hopefully, it will be
possible to measure this state in the upcoming studies of GlueX and CLAS12 at
Jefferson lab.\ We comment further on this possibility in the conclusions.

(v) Radiative decays. The results are reported in Table 11. Experimentally,
they were not yet seen. The magnitude of these decay widths is similar to the
one of the lighter nonet of excited vector mesons, compare with Table 8. In
particular, the largest decay rate is $\omega(1650)\rightarrow\gamma\pi$, in
agreement with the fact that $\omega(1650)\rightarrow\rho\pi$ is dominant. In
general, these radiative decays seem quite interesting and important for the
future studies of this nonet.

\bigskip

Finally, the nonet \{$\rho(1700)$, $K^{\ast}(1680)$, $\omega(1650)$,
$\phi(???)\equiv\phi(1930)$\} is well compatible with a nonet of excited
vector mesons, predominantly corresponding to orbitally excited vector states.
However, the errors of the theoretical results are quite large and some
experimental results are not yet fully in agreement with each other. Hence,
even if the qualitative picture is quite satisfactory, there is room for
quantitative improvements. Moreover, the experimental determination of
radiative decays and the measurement of the yet missing state $\phi(1930)$
represent a useful test to fully establish the nature of this nonet.

\section{Discussions and Conclusions}

In this work we have studied the strong and radiative decays of the vector
mesons \{$\rho(1450),$ $K^{\ast}(1410),$ $\omega(1420),$ $\phi(1680)$\} and
\{$\rho(1700),$ $K^{\ast}(1680),$ $\omega(1650),$ $\phi(???)\equiv\phi
(1930)$\} by using a flavour-invariant QFT Lagrangian approach. This
Lagrangian contains 4 coupling constants, corresponding to the dominant
interaction terms in the large-$N_{c}$ expansion, have been determined by
using four well-known experimental quantities. Then, we have compared our
results to the averages and fits of the PDG as well as to selected experiments
listed therein, see Tables 6-11. Moreover, we have studied a large number of
ratios for which an experimental counterpart was measured or could be deduced
by combining present data. In summary, the assignment of these mesonic states
to (predominantly) radially excited and to orbitally excited vector mesons
works well. Typically, the dominant decays seen in experiment are also the
leading ones in theory, while those decays which were not yet seen in
experiment are generally quite small theoretically. In the future, it will be
possible to further test our theoretical approach by measuring those decays
which are not yet listed in PDG. In some cases, some decay ratios which were
measured by more than one experiments, are not in agreement with each other.
Also along this direction, future determinations will be useful.

Besides strong decays, we have also calculated radiative decays of the type
$R\rightarrow\gamma P$ by using vector meson dominance. For the lighter nonet
\{$\rho(1450),$ $K^{\ast}(1410),$ $\omega(1420),$ $\phi(1680)$\} some
radiative decays have been measured (in a couple of cases even the
corresponding decay width can be determined from existing data); for the
heavier nonet $\{\rho(1700),K^{\ast}(1680),\omega(1650),\phi(???)\}$, no
experimental results exists at present. Hence, our results are predictions.
The study of radiative decays of excited vector mesons seems quite promising
in future experimental activities.

One important outcome of our approach concerns the predictions for a novel
$\phi$ state, belonging to the heavier nonet (predominantly orbitally excited
vector mesons). By comparison with the mass differences between the two
nonets, we have estimated that the mass is about $1930$ MeV, hence we have
called this state $\phi(1930).$ This mass is not far from the quark model
prediction of $1890$ MeV \cite{godfrey} and is also compatible with the
lattice result of Ref. \cite{dudek} in which the mass of this predominantly
$\bar{s}s$ state is about $1950$ MeV. In Table 12 we summarize the results for
the putative state $\phi(1930).$ Note, the $KK$ decay is about $100$ MeV
\cite{godfrey}, which turns out to be similar to the results of the quark
model. On the contrary, the mode $\phi(1930)\rightarrow K^{\ast}(892)K$ is
quite different: the quark model predicts $50$ MeV, sizably smaller than the
our result reported in Table 12 (even if the errors are large, one can
conclude that in our approach the corresponding partial decay width is larger
than $100$ MeV. In general, we predict that $\Gamma_{\phi(1930)\rightarrow
K^{\ast}K}>\Gamma_{\phi(1930)\rightarrow KK}$)$.$

\begin{center}
\begin{table}[h]
\makebox[\textwidth][c] {%
\begin{tabular}
[c]{|c|c|}\hline
\multicolumn{2}{|c|}{MESON $\phi(1930)$}\\\hline\hline
Quark composition & $\approx s\bar{s}$\\\hline
Old spectroscopy notation & (predom.) $n\hspace{0.08cm}^{2S+1}L_{J}=1^{3}%
D_{1}$\\\hline
$n$ & (predom.) $1$\\\hline
$S$ & (predom.) $1$\text{\quad}$\uparrow\uparrow$\\\hline
$L$ & (predom.) $2$\\\hline
$J^{PC}$ & $1^{--}$\\\hline
Mass & $\approx1930\pm40$ MeV\\\hline\hline
\multicolumn{2}{|c|}{DECAYS}\\\hline
Decay channel & Decay width\\
& [MeV]\\\hline
$\phi(1930)\rightarrow\bar{K}K$ & $104\pm28$\\\hline
$\phi(1930)\rightarrow K\bar{K}^{\ast}$ & $260\pm109$\\\hline
$\phi(1930)\rightarrow\Phi(1020)\eta$ & $67\pm28$\\\hline
$\phi(1930)\rightarrow\Phi(1020)\eta^{\prime}$ & $\approx0$\\\hline
$\phi(1930)\rightarrow\gamma\eta$ & $0.19\pm0.12$\\\hline
$\phi(1930)\rightarrow\gamma\eta^{\prime}$ & $0.13\pm0.08$\\\hline
\end{tabular}
}\caption{Summary table for the putative state $\phi(1930).$}%
\end{table}
\end{center}

The putative state $\phi(1930)$ is quite broad, thus making its discovery more
difficult. Yet, a dedicated search by using partial wave analysis could reveal
the existence of this state. In general, the very promising GlueX
\cite{gluex,gluex2,gluex3} and CLAS12 \cite{clas12} experiments take place in
the near future. The process in\ Fig. 1%
\begin{equation}
\gamma+p\rightarrow K^{+}+K^{-}+p\text{ , }\gamma+p\rightarrow K^{0}+\bar
{K}^{0}+p\label{photoproduction}%
\end{equation}
is an example of a process that can be studied at GlueX and CLAS12. Quite
remarkably, each mesonic vertex is contained in the present paper:
$\phi(1930)\gamma\eta$ and $\phi(1930)KK$. An important outlook of the present
work is a dedicated study of this reaction. For the baryonic part, one can use
a well defined hadronic model containing baryons and their interactions with
mesons (in particular, the coupling of the nucleons to the $\eta$ meson is
necessary), as for instance the extended Linear Sigma Model (eLSM) based on
the mirror assignment presented in\ Refs. \cite{gallas,olbrich1,olbrich2}. The
analogous diagram in which $\phi(1930)$ couples to $K^{\ast}(892)K$
\begin{equation}
\gamma+p\rightarrow K^{\ast+}(892)+K^{-}+p\rightarrow K^{+}+K^{-}+\pi
^{0}+p\text{ ,}%
\end{equation}
also takes place (together with analogous isospin related reactions) and
should be studied in the same context.

\begin{figure}[h]
\begin{center}
\includegraphics[width=0.5 \textwidth] {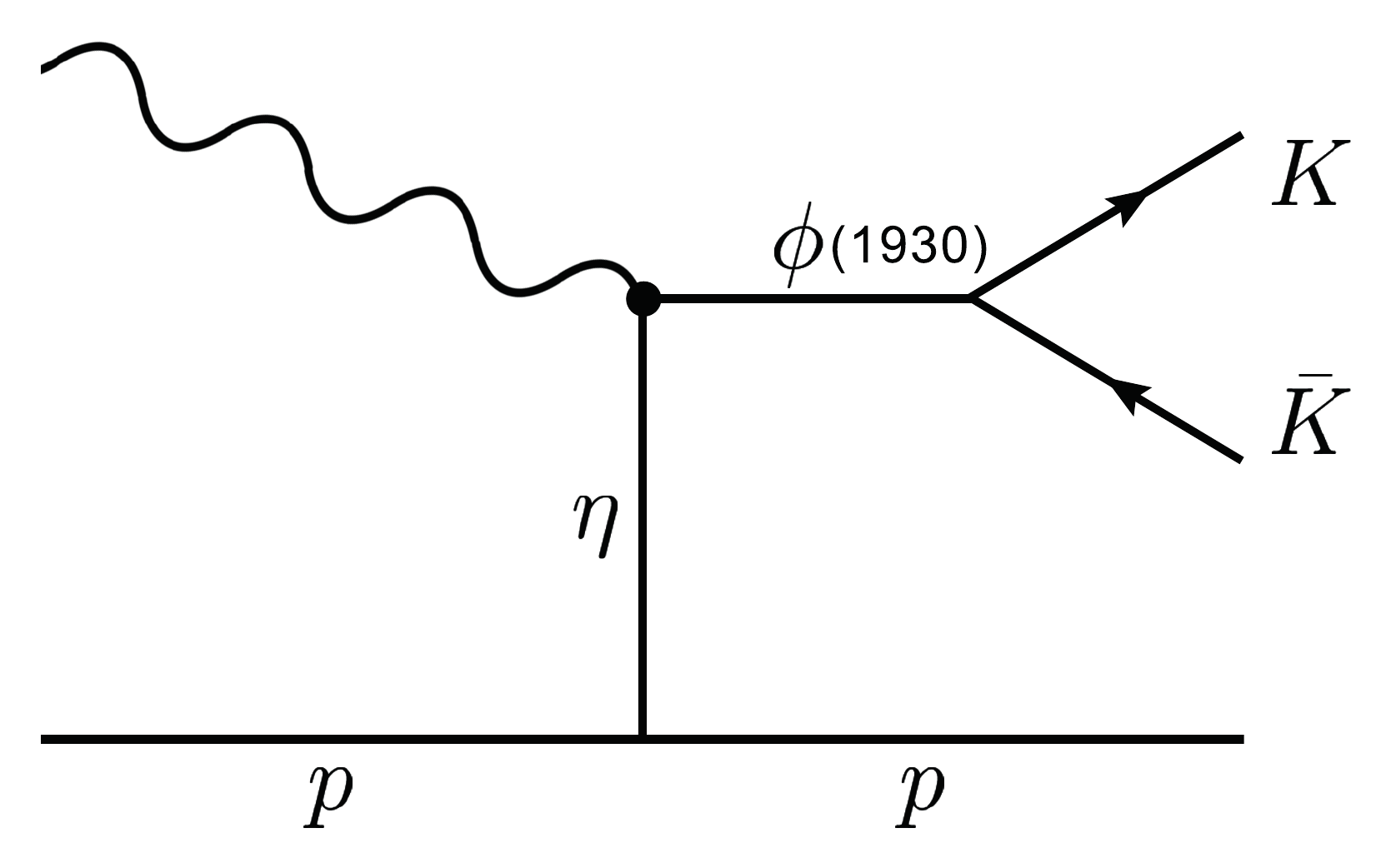}
\end{center}
\caption{Feynman diagram for the process of Eq. (\ref{photoproduction}).}%
\end{figure}

Other experiments are important as well. In principle, the resonance
$\phi(1930)$ should also be contained in the data of BABAR \cite{babar}, in
which the reaction $e^{+}e^{-}\rightarrow K^{+}K^{-}$ was studied. Namely, in
this reaction all the vector mesons $\rho^{0}(1450),$ $\omega(1420),$
$\phi(1680),$ $\rho^{0}(1700),\omega(1650),\phi(???)\equiv\phi(1930)$ enter
and an interference of different amplitudes takes place (for a recent
analysis, see Ref. \cite{vladimir} and refs. therein). Similarly, BESIII can
also study a similar reaction, but typically this experiment focussed on the
range of energy above 2 GeV.

In the future, PANDA will be a leading experiment for spectroscopy
\cite{panda}.\ While the energy in the center of mass will be too high to
create excited vector mesons in a fusion process, it will be well possible to
produce excited vector mesons together with light mesons (such as pions and kaons).

On the theoretical side, we enumerated the possible straightforward
improvements of our approach: the systematic inclusion of large-$N_{c}$ and
flavour-symmetry violating terms. With new and more precise data, this study
can be easily performed. A further outlook consists in the extension of the
model by using chiral symmetry. Here, it was not needed because we did not
link the excited vector mesons to chiral partners. For instance, the chiral
partners of the orbitally excited vector mesons are the rather well-known
pseudovector mesons \{$b_{1}(1235),$ $K_{1,B},$ $h_{1}(1170),$ $h_{1}(1380)$\}
(for the corresponding mathematical set-up, see Ref. \cite{vg}). Chiral
symmetry can help to relate the decays of this nonet to the decays of
\{$\rho(1700),$ $K^{\ast}(1680),$ $\omega(1650),$ $\phi(???)\equiv\phi
(1930)$\}. The same extension to radially excited vector mesons is more
difficult because the corresponding chiral partners, a nonet of excited
axial-vector states, has not been yet experimentally discovered.

In conclusions, the theoretical and experimental study of the two nonets of
excited vector mesons is an interesting subject of low-energy spectroscopy.
While the qualitative picture seems clear, further measurements, with special
attention on radiative decays, are needed to fully establish the nature of
these states. Moreover, the discovery of $\phi(1930)$ would represent a nice
confirmation of the quark-antiquark picture also for excited states in the
low-energy domain.

\section*{Acknowledgements}

The authors thank S.\ Coito, V. Sauli, J. Sammet, D. H. Rischke, D.
Parganlija, and Peter Kovacs for useful discussions. M.P. and F.G. acknowledge
support from the Polish National Science Centre (NCN) through the OPUS project
no. 2015/17/B/ST2/01625.

\appendix

\section{Extended form of the Lagrangian}

In this Appendix we present the extended expressions of the Lagrangian
introduced in Sec. 2.2, Eq. (\ref{lagfull}).

The Lagrangian terms of our model $\mathcal{L}_{iPP}$ with $i=E,D$ presented
in Sec. 2, Eq. (\ref{lagfull1}), are given by:
\begin{equation}
\begin{aligned} \mathcal{L}_{iPP} &= i g_{iPP}\; \text{Tr}\big[[\partial^\mu P, V_{i,\mu}]P\big] \\ &= \frac{ig_{iPP}}{4}\; \Bigl\{ {K^*_\mu}^0 \Big( (\partial^\mu \overline{K}^0) \pi^0 - \overline{K}^0 (\partial^\mu \pi^0) - \sqrt{2}(\partial^\mu K^-) \pi^+ + \sqrt{2} K^- (\partial^\mu \pi^+)\\ &+ (\partial^\mu \eta_N) \overline{K}^0 - \eta_N (\partial^\mu \overline{K}^0) + \sqrt{2} \eta_S (\partial^\mu \overline{K}^0) - \sqrt{2} (\partial^\mu \eta_S) \overline{K}^0 \Big)\\ &+ {\overline{K}^*_\mu}^0 \Big( K^0 (\partial^\mu\pi^0) - (\partial^\mu K^0) \pi^0 - \sqrt{2} K^+ (\partial^\mu \pi^-) + \sqrt{2} (\partial^\mu K^+) \pi^- \\ &+ \eta_N (\partial^\mu K^0) - (\partial^\mu \eta_N) K^0 - \sqrt{2} \eta_S (\partial^\mu K^0) + \sqrt{2} (\partial^\mu \eta_S) K^0\Big)\\ &+ {K^*_\mu}^- \Big( (\partial^\mu K^+) \pi^0 - K^+ (\partial^\mu \pi^0) - \sqrt{2} K^0 (\partial^\mu \pi^+) + \sqrt{2} (\partial^\mu K^0) \pi^+\\ &+ \eta_N (\partial^\mu K^+) - (\partial^\mu \eta_N) K^+ - \sqrt{2}\eta_S (\partial^\mu K^+) + \sqrt{2}(\partial^\mu \eta_S) K^+ \Big)\\ &+ {K^*_\mu}^+ \Big( K^- (\partial^\mu \pi^0) - (\partial^\mu K^-) \pi^0 - \sqrt{2} (\partial^\mu \overline{K}^0) \pi^- + \sqrt{2} \overline{K}^0 (\partial^\mu \pi^-) \\ &+ (\partial^\mu \eta_N) K^- - \eta_N (\partial^\mu K^-) + \sqrt{2}\eta_S (\partial^\mu K^-) - \sqrt{2} (\partial^\mu \eta_S) K^-\Big) \\ &+ \rho^0_\mu \Big( \overline{K}^0 (\partial^\mu K^0) - (\partial^\mu \overline{K}^0) K^0 + K^+ (\partial^\mu K^-) - (\partial^\mu K^+) K^- + 2\pi^+ (\partial^\mu \pi^-) - 2(\partial^\mu \pi^+) \pi^-\Big) \\ &+ \rho^-_\mu \Big( \sqrt{2} K^+ (\partial^\mu \overline{K}^0) - \sqrt{2}(\partial^\mu K^+) \overline{K}^0 +2\pi^0 (\partial^\mu \pi^+) - 2 (\partial^\mu \pi^0) \pi^+ \Big) \\ &+ \rho^+_\mu \Big( \sqrt{2} K^0 (\partial^\mu K^-) - \sqrt{2}(\partial^\mu K^0) K^- + 2(\partial^\mu \pi^0) \pi^- - 2\pi^0 (\partial^\mu \pi^-)\Big) \\ &+ \omega \Big( K^0 (\partial^\mu \overline{K}^0) - (\partial^\mu K^0) \overline{K}^0 + K^+ (\partial^\mu K^-) - (\partial^\mu K^+) K^- \Big) \\ &+ \sqrt{2}\phi \Big( (\partial^\mu K^0) \overline{K}^0 - K^0 (\partial^\mu \overline{K}^0) - K^+ (\partial^\mu K^-) + (\partial^\mu K^+) K^- \Big) \Bigr\}. \end{aligned}
\end{equation}
We recall that for $i=E$ the states correspond to $\{\rho,K^{\ast},\phi
,\omega\}=$\{$\rho(1450),$ $K^{\ast}(1410),$ $\omega(1420),$ $\phi(1680)$\}
and for $i=D$ to $\{\rho,K^{\ast},\phi,\omega\}=$\{$\rho(1700),$ $K^{\ast
}(1680),$ $\omega(1650)$, $\phi(1930)$\}.

The Lagrangian terms of our model $\mathcal{L}_{iVP}$ with $i=E,D$ presented
in Sec. 2, Eq. (\ref{lagfull2}), are given by:%

\begin{equation}
\begin{aligned} \mathcal{L}_{iVP} &= g_{iVP}\; \text{Tr}\big(\tilde{V}^{\mu\nu}_{i}\lbrace V_{\mu\nu},P\rbrace\big) = 2g_{iVP}\; \epsilon^{\mu\nu\alpha\beta}\; \text{Tr}\big( (\partial_\alpha V_{i,\beta}) \{ (\partial_\mu V_{\nu}), P \} \big)\\ &= \frac{g_{iVP}}{2}\epsilon^{\mu\nu\alpha\beta}\; \Bigl\{ (\partial_\alpha \rho_{i,\beta}^0) \Big( 2\pi^0 (\partial_\mu \omega_\nu) + 2\eta_N (\partial_\mu \rho_\nu^0) - \overline{K}^0 (\partial_\mu {K^*_\nu}^0) - K^0 (\partial_\mu {\overline{K}^*_\nu}^0) + K^+(\partial_\mu {K^*_\nu}^-) + K^- (\partial_\mu {K^*_\nu}^+) \Big)\\ &+ \sqrt{2}(\partial_\alpha \rho_{i,\beta}^-) \Big( \sqrt{2}\pi^+ (\partial_\mu \omega_\nu) + \sqrt{2}\eta_N (\partial_\mu \rho^+_\nu) + K^+(\partial_\mu {\overline{K}^*_\nu}^0) +\overline{K}^0 (\partial_\mu {K^*_\nu}^+) \Big)\\ &+ \sqrt{2}(\partial_\alpha \rho_{i,\beta}^+) \Big( \sqrt{2}\pi^- (\partial_\mu \omega_\nu) + \sqrt{2} \eta_N (\partial_\mu \rho^-_\nu) + K^- (\partial_\mu {K^*_\nu}^0) + K^0 (\partial_\mu {K^*_\nu}^-) \Big)\\ &+ \sqrt{2}(\partial_\alpha \phi_{i,\beta}) \Big( 2\eta_S (\partial_\mu \phi_\nu) + K^0 (\partial_\mu {\overline{K}^*_\nu}^0) + \overline{K}^0 (\partial_\mu {K^*_\nu}^0) + K^+ (\partial_\mu {K^*_\nu}^-) + K^- (\partial_\mu {K^*_\nu}^+) \Big) \\ &+ (\partial_\alpha \omega_{i,\beta}) \Big( 2\pi^0 (\partial_\mu \rho^0_\nu) + 2\pi^+ (\partial_\mu \rho^-_\nu) + 2\pi^- (\partial_\mu \rho^+_\nu) + 2\eta_N (\partial_\mu \omega_\nu)\\ &+ K^0 (\partial_\mu {\overline{K}^*_\nu}^0) + \overline{K}^0 (\partial_\mu {K^*_\nu}^0) + K^+ (\partial_\mu {K^*_\nu}^-) + K^- (\partial_\mu {K^*_\nu}^+)\Big)\\ &+ (\partial_\alpha K^{*0}_{i,\beta}) \Big( \overline{K}^0 (\partial_\mu \omega_\nu) - \pi^0 (\partial_\mu {\overline{K}^*_\nu}^0) + \sqrt{2} \pi^+ (\partial_\mu {K^*_\nu}^-) - \overline{K}^0 (\partial_\mu \rho^0_\nu) + \sqrt{2}K^- (\partial_\mu \rho^+_\nu) \\ &+ \eta_N (\partial_\mu {\overline{K}^*_\nu}^0) + \sqrt{2}\eta_S (\partial_\mu {\overline{K}^*_\nu}^0) + \sqrt{2}\overline{K}^0 (\partial_\mu \phi_\nu) \Big) \\ &+ (\partial_\alpha \overline{K}^{*0}_{i,\beta}) \Big( K^0 (\partial_\mu \omega_\nu) - \pi^0 (\partial_\mu {K^*_\nu}^0) + \sqrt{2}\pi^- (\partial_\mu {K^*_\nu}^+) - K^0 (\partial_\mu \rho^0_\nu) + \sqrt{2}K^+ (\partial_\mu \rho^-_\nu) \\ &+ \eta_N (\partial_\mu {K^*_\nu}^0) + \sqrt{2}\eta_S (\partial_\mu {K^*_\nu}^0) + \sqrt{2} K^0 (\partial_\mu \phi_\nu) \Big) \\ &+ (\partial_\alpha K^{*-}_{i,\beta}) \Big( K^+ (\partial_\mu \omega_\nu) + \pi^0 (\partial_\mu {K^*_\nu}^+)+ \sqrt{2}\pi^+ (\partial_\mu {K^*_\nu}^0) + K^+ (\partial_\mu \rho^0_\nu) + \sqrt{2}K^0 (\partial_\mu \rho^+_\nu) \\ &+ \eta_N (\partial_\mu {K^*_\nu}^+) +\sqrt{2}\eta_S (\partial_\mu {K^*_\nu}^+) + \sqrt{2} K^+ (\partial_\mu \phi_\nu) \Big) \\ &+ (\partial_\alpha K^{*+}_{i,\beta}) \Big( K^- (\partial_\mu \omega_\nu) + \pi^0 (\partial_\mu {K^*_\nu}^-) + \sqrt{2}\pi^- (\partial_\mu {\overline{K}^*_\nu}^0) + K^- (\partial_\mu \rho^0_\nu) + \sqrt{2}\overline{K}^0 (\partial_\mu \rho^-_\nu) \\ &+ \eta_N (\partial_\mu {K^*_\nu}^-) + \sqrt{2}\eta_S (\partial_\mu {K^*_\nu}^-) + \sqrt{2} K^- (\partial_\mu \phi_\nu) \Big) \Bigr\}. \end{aligned}
\end{equation}
We recall that for $i=E$ the states correspond to $\{\rho,K^{\ast},\phi
,\omega\}=$\{$\rho(1450),$ $K^{\ast}(1410),$ $\omega(1420),$ $\phi(1680)$\}
and for $i=D$ to $\{\rho,K^{\ast},\phi,\omega\}=$\{$\rho(1700),$ $K^{\ast
}(1680),$ $\omega(1650)$, $\phi(1930)$\}.

\section{Coupling to the photon via VMD}

Let us start from a single neutral $\rho^{0}$ meson. Its coupling to an
electron-positron pair can be written down as:
\begin{equation}
\mathcal{L}_{\rho e^{+}e^{-}}=g_{\rho e^{+}e^{-}}\rho_{\mu}^{0}\bar{\psi}%
_{e}\gamma^{\mu}\psi_{e}\text{ .} \label{rhoepem}%
\end{equation}
Then, the decay into $e^{+}e^{-}$ reads:
\begin{equation}
\Gamma_{\rho\rightarrow e^{+}e^{-}}=\frac{\sqrt{\frac{m_{\rho}^{2}}{4}%
-m_{e}^{2}}}{6\pi m_{\rho}^{2}}\left(  m_{\rho}^{2}+2m_{e}^{2}\right)  g_{\rho
e^{+}e^{-}}^{2}\text{.}%
\end{equation}
The interaction (\ref{rhoepem}) can be obtained in the framework of Vector
Meson Dominance (according to the so-called VMD-1 of Ref. \cite{connell}) by
considering from the Lagrangian%
\begin{equation}
\mathcal{L}_{VMD,\rho}=e_{0}A_{\mu}\bar{\psi}_{e}\gamma^{\mu}\psi_{e}%
-\frac{e_{0}}{2g_{\rho}}\rho_{\mu\nu}^{0}F^{\mu\nu}\text{ ,}%
\end{equation}
where the coupling $g_{\rho}$ appears also in the decay amplitude of the
process $\rho^{0}\rightarrow\pi^{+}\pi^{-}$. The $\rho^{0}$ meson first
transforms to a photon, which then generates a lepton pair. As a consequence,
the following relation holds:
\begin{equation}
g_{\rho e^{+}e^{-}}=\frac{e_{0}}{g_{\rho}}%
\end{equation}
In fact $\rho_{\mu\nu}^{0}F^{\mu\nu}\rightarrow2q^{2}\rho_{\mu}^{0}A^{\mu},$
hence one gets in the corresponding amplitude (upon using the Feynman rules):
\begin{equation}
\frac{e_{0}}{2g_{\rho}}2q^{2}\frac{1}{q^{2}}=\frac{e_{0}}{g_{\rho}}.
\end{equation}
Finally, VMD-1 implies that:
\begin{equation}
\Gamma_{\rho\rightarrow e^{+}e^{-}}=\frac{\sqrt{\frac{m_{\rho}^{2}}{4}%
-m_{e}^{2}}}{6\pi m_{\rho}^{2}}\left(  m_{\rho}^{2}+2m_{e}^{2}\right)  \left(
\frac{e}{g_{\rho}}\right)  ^{2}.
\end{equation}
The very same formula can be used for the decay into a muon pair:
\begin{equation}
\Gamma_{\rho\rightarrow\mu^{+}\mu^{-}}=\frac{\sqrt{\frac{m_{\rho}^{2}}%
{4}-m_{\mu}^{2}}}{6\pi m_{\rho}^{2}}\left(  m_{\rho}^{2}+2m_{\mu}^{2}\right)
\left(  \frac{e}{g_{\rho}}\right)  ^{2}.
\end{equation}
Moreover, also the decay of the other neutral scalar states can be obtained
(straightforward changes due to different charges of quarks must be taken into
account):
\begin{align}
\Gamma_{\omega\rightarrow e^{+}e^{-}}  &  =\frac{\sqrt{\frac{m_{\omega}^{2}%
}{4}-m_{e}^{2}}}{6\pi m_{\omega}^{2}}\left(  m_{\omega}^{2}+2m_{e}^{2}\right)
\left(  \frac{e}{3g_{\rho}}\right)  ^{2}\text{ ,}\\
\Gamma_{\phi\rightarrow e^{+}e^{-}}  &  =\frac{\sqrt{\frac{m_{\phi}^{2}}%
{4}-m_{e}^{2}}}{6\pi m_{\phi}^{2}}\left(  m_{\phi}^{2}+2m_{e}^{2}\right)
\left(  -\frac{\sqrt{2}}{3}\frac{e}{g_{\rho}}\right)  ^{2}.
\end{align}
The extension of the latter two to the decays into a muon pair is straightforward.

The extension to the full nonet is then obtained by using the matrix for
vector mesons introduced in\ Sec. 2, which we rewrite here for convenience:%
\begin{equation}
V_{\mu}=\frac{1}{\sqrt{2}}\left(
\begin{array}
[c]{ccc}%
\frac{\omega}{\sqrt{2}}+\frac{\rho^{0}}{\sqrt{2}} & \rho^{+} & K(892)^{\ast
+}\\
\rho^{-} & \frac{\omega}{\sqrt{2}}-\frac{\rho^{0}}{\sqrt{2}} & K(892)^{\ast
0}\\
K(892)^{\ast-} & \bar{K}(892)^{\ast0} & \phi
\end{array}
\right)  \text{ .}%
\end{equation}
The VMD approach reads:%
\begin{equation}
\mathcal{L}_{VMD,full}=e_{0}A_{\mu}\bar{\psi}_{e}\gamma^{\mu}\psi_{e}-\frac
{e}{g_{\rho}}F_{\mu\nu}Tr[V^{\mu\nu}Q]
\end{equation}
with%
\begin{equation}
Q=\left(
\begin{array}
[c]{ccc}%
2/3 & 0 & 0\\
0 & -1/3 & 0\\
0 & 0 & -1/3
\end{array}
\right)
\end{equation}
Finally, the photon-meson mixing can be taken into account by performing the
shift:
\begin{equation}
V_{\mu\nu}\rightarrow V_{\mu\nu}+\frac{e_{0}}{g_{\rho}}F_{\mu\nu}Q.
\end{equation}
This is the shift that we have applied in order to determine the decay of the
type $R\rightarrow\gamma P$ studied in this work. Intuitively, one has a decay
chain of the type $R\rightarrow VP\rightarrow\gamma P,$ where in the second
step the transition $V\rightarrow\gamma$ has taken place according to VMD.

\section{Errors of the coupling constants and their propagation}

Let us consider the $\chi^{2}$ function  $F\equiv F(x_{k}),$ where $x_{k}$ are
the parameters of the theory with $k=1,...,N$ (in our cases, $F$ corresponds
to Eq. (\ref{fe}) and to Eq. (\ref{fd}) respectively, and the parameters
$x_{1}$ and $x_{2}$ are the coupling constants). We look for the minimum of
$F$ by solving $\partial_{q}F(x_{k})=0\rightarrow x_{k}=x_{k}^{\min}.$ The
Taylor expansion reads%
\begin{equation}
F(x_{k})=F(x_{k}^{\min})+(x_{k}-x_{k}^{\min})H_{kq}(x_{q}-x_{q}^{\min
})+...\text{, }H_{kq}=\left. \frac{1}{2} \frac{\partial^{2}F}{\partial x_{k}\partial
x_{q}}\right\vert _{x_{k}=x_{k}^{\min}}%
\end{equation}
The matrix $H$, with elements $H_{kq}$, is the Hesse matrix of the function
$F$ evaluated at the minimum. We introduce the matrix $B$ such that
$BHB^{t}=D=diag\{\lambda_{1},...,\lambda_{N}\}$ and the new variables
$z_{k}=B_{kq}(x_{q}-x_{q}^{\min})$ (note: $B_{kq}=\frac{\partial z_{k}%
}{\partial x_{q}}$). As function of $z_{k}$, we get
\begin{equation}
F\equiv F(z_{k})=F(x_{k}^{\min})+z_{k}^{2}\lambda_{k}+...\text{ ,}%
\end{equation}
therefore the error of $z_{k}$ is given by $\delta z_{k}=1/\sqrt{\lambda_{k}}$
(increment of $1$ of the $\chi^{2}$). Next, let us consider an arbitrary
function of the parameters, $G\equiv G(x_{k})$, which represents some physical
quantity of interest (in our examples, it can be a decay width or a ratio of
decay widths). The physical value of $G$ is clearly given by $G(x_{k}^{\min
}).$ Its error is evaluated w.r.t. the (mutually independent) parameters $z_{k}$:
\begin{equation}
\delta G=\sqrt{\left(  \left.  \frac{\partial G}{\partial z_{k}}\right\vert
_{z_{k}=0}\delta z_{k}\right)  ^{2}}\text{ ,}%
\end{equation}
where
\begin{equation}
\left.  \frac{\partial G}{\partial z_{k}}\right\vert _{z_{k}=0}=\left.
\frac{\partial G}{\partial x_{q}}\right\vert _{x_{k}=x_{k}^{\min}}B_{qk}%
^{t}=\left.  \frac{\partial G}{\partial x_{q}}\right\vert _{x_{k}=x_{k}^{\min
}}B_{kq}\text{ .}%
\end{equation}
The errors of the parameters $x_{r}$ is calculated by setting $G=x_{r},$ out
of which (upon using $\frac{\partial x_{r}}{\partial z_{k}}=B_{rk}^{t}=B_{kr}%
$):%
\begin{equation}
\delta x_{r}=\sqrt{\left(  B_{kr}\delta z_{k}\right)  ^{2}}=\sqrt{H_{rr}^{-1}%
}.
\end{equation}
(For the last equality we used $H^{-1}=B^{t}D^{-1}B\rightarrow H_{rr}%
^{-1}=B_{rq}^{t}D_{qk}^{-1}B_{kr}=B_{rk}^{t}\delta z_{k}^{2}B_{kr}=B_{kr}%
^{2}\delta z_{k}^{2}$). In this way we evaluated the parameter errors in Sec.
3.1. It is also interesting to mention that the naive evaluation of the error
of $G$ as
\begin{equation}
\delta G^{naive}=\sqrt{\left(  \left.  \frac{\partial G}{\partial x_{k}%
}\right\vert _{x_{k}=x_{k}^{\min}}\delta x_{k}\right)  ^{2}}%
\end{equation}
is in general not correct (typically, it is an overestimation of the error
$\delta G$).

Finally, we turn to our concrete examples. For the nonet of radially excited
vector states, we identify $x_{1}=g_{EPP}$ and $x_{2}=g_{EVP},$ and $F$ is
given by Eq. (\ref{fe}).\ Here, the Hesse matrix is from the very beginning
diagonal. Than, in this particular case $\delta G^{naive}=\delta G,$ hence the
error of a certain quantity $G(g_{EPP}$,$g_{EVP})$ reads:
\begin{equation}
\delta G=\sqrt{\left(  \left.  \frac{\partial G}{\partial g_{EPP}}\right\vert
_{\min}\delta g_{EPP}\right)  ^{2}+\left(  \left.  \frac{\partial G}{\partial
g_{EVP}}\right\vert _{\min}\delta g_{EVP}\right)  ^{2}}%
\end{equation}
where `min' refers to the values of Eq. (\ref{ge}). In this way all the errors
of the quantities of Sec. 3.2 were evaluated.

For what concerns the nonet of orbitally excited vector states, we set
$x_{1}=g_{DPP}$ and $x_{2}=g_{DVP}$ and use $F$ from Eq. (\ref{fd}). While for
the decays in\ Tables 9, 10, and 11 (which involve only one of the two coupling
constants) the naive procedure would still be valid, this is not true in
general and the diagonalization is necessary. For instance, for the ratios of
coupling constants (entering in various decay ratios studied in\ Sec. 3.3),
one sets $G=g_{DPP}^{2}/g_{DVP}^{2}$ . The central value reads $5.4$ and the
corresponding error is $\delta G=1.8$, whereas $\delta G^{naive}=2.7$ would be
an overestimation.

In this work, we have limited our evaluation of the errors to the couplings
discussed above because they represent the largest source of inderterminacy.
Yet, as mentioned in the text, other error sources (not included here) exist,
such as masses and flavour-breaking and large-$N_{c}$ suppressed terms.

\end{document}